%% file: ds50veto_electronics.tex
\pdfoutput=1 
\documentclass{JINST}

\input{def.tex}

\title{The Electronics and Data Acquisition System for the DarkSide-50 Veto Detectors}
\input{authors.tex}

\abstract{DarkSide-50 is a detector for dark matter candidates in the form of weakly interacting massive particles (WIMPs). 
It utilizes a liquid argon time projection chamber (LAr TPC) for the inner main detector.
The TPC is surrounded by a liquid scintillator veto (LSV) and a water Cherenkov veto detector (WCV).
The LSV and WCV, both instrumented with PMTs, act as the neutron and cosmogenic muon veto detectors for DarkSide-50.
This paper describes the electronics and data acquisition system used for these two detectors.}


\begin{document}

\input{ds50experiment.tex}

\input{ds50vetoes.tex}

\input{vetorequirements.tex}

\input{vetofe.tex}

\input{vetodaq.tex}

\input{performances.tex}

\input{conclusions.tex}

\acknowledgments

The DarkSide-50 Collaboration would like to thank LNGS laboratory and its staff for invaluable technical and logistical support.  This report is based upon work supported by the US NSF (Grants PHY-0919363, {PHY}-{1004072}, {PHY}-{1004054}, {PHY}-{1242585}, {PHY}-{1314483}, {PHY}-{1314507} and associated collaborative grants; Grants {PHY}-{1211308} and {PHY}-{1455351}), the Italian Istituto Nazionale di Fisica Nucleare (INFN), the US DOE (Contract Nos.~{DE}-{FG02-91ER40671} and {DE}-{AC02-07CH11359}), the Polish NCN (Grant {UMO}-{2012/05/E/ST2/02333}), and the Russian Science Foundation Grant No 16-12-10369. We thank the staff of the Fermilab
Particle Physics, Scientific and Core Computing Divisions for their support.  We acknowledge the financial support from the UnivEarthS Labex program of Sorbonne Paris Cit\'e ({ANR}-{10-LABX-0023} and {ANR}-{11-IDEX-0005-02}) and from the S\~ao Paulo Research Foundation (FAPESP).

The collaboration gratefully acknowledges the support of  Nicola Cescato, Fabio Cortinovis, Thierry Debelle, and Andrea Nobile of National Instruments.

\input{bibliography.tex}
\end{document}

%% file: def.tex
\usepackage{comment}
\usepackage{url}

\newcommand{\lith}{\mbox{$^{7}$Li}}

\newcommand{\borten}{\mbox{$^{10}$B}}

\newcommand{\cfor}{\mbox{$^{14}$C}}

\newcommand{\ar}{\mbox{$^{39}$Ar}}

\newcommand{\coba}{\mbox{$^{60}$Co}}
\newcommand{\tmbchem}{\mbox{B(OCH$_3$)$_3$}}

\newcommand{\daq}{\mbox{DAQ}}
\newcommand{\wimp}{\mbox{WIMP}}
\newcommand{\wimps}{\mbox{WIMPs}}

\newcommand{\dsf}{\mbox{DarkSide-50}}
\newcommand{\bx}{\mbox{Borexino}}

\newcommand{\lsv}{\mbox{LSV}}
\newcommand{\wcd}{\mbox{WCV}}

\newcommand{\tpc}{\mbox{TPC}}
\newcommand{\pmt}{\mbox{PMT}}
\newcommand{\pmts}{\mbox{PMTs}}

\newcommand{\tmb}{\mbox{TMB}}
\newcommand{\pc}{\mbox{PC}}

\newcommand{\aar}{\mbox{AAr}}
\newcommand{\lar}{\mbox{LAr}}

\newcommand{\uar}{\mbox{UAr}}

\newcommand{\gr}{\mbox{$\gamma$-ray}}
\newcommand{\grs}{\mbox{$\gamma$-rays}}

\newcommand{\lsvpmtnum}{\mbox{110}}

\newcommand{\lsvpmt}{\mbox{Hamamatsu R5912}}
\newcommand{\lsvpmtsize}{\mbox{8}''}
\newcommand{\lsvpmtqe}{\mbox{37}\,\%}
\newcommand{\lsvpmtwave}{\mbox{408}\,{nm}}
\newcommand{\lsvcforrate}{$\sim$150\,kBq}      

\newcommand{\lsvbtenxsec}{\mbox{3840}\,\mbox{barn}}
\newcommand{\ctfpmtnum}{\mbox{80}}

\newcommand{\ctfpmt}{\mbox{ETL 9351}}
\newcommand{\ctfpmtsize}{\mbox{8}{''}}
\newcommand{\ctfpmtqe}{\mbox{27}\,{\%}}
\newcommand{\ctfpmtwave}{\mbox{420}\,{nm}}

\newcommand{\feab}{\mbox{FEAB}}
\newcommand{\feabs}{\mbox{FEABs}}
\newcommand{\fedb}{\mbox{FEDB}}
\newcommand{\fedbs}{\mbox{FEDBs}}

\newcommand{\abborten}{\mbox{20}\,\%}

\newcommand{\dsfexpoaar}{\mbox{1422(67)}\,kg\,day}
\newcommand{\dsfexpouar}{\mbox{2616(67)}\,kg\,day}
\newcommand{\pxie}{\mbox{PXIe}}
\newcommand{\pxiechassis}{NI \pxie-1075}
\newcommand{\pxiecontroller}{NI \pxie-8133}
\newcommand{\pxiedigitizer}{NI \pxie-5162}
\newcommand{\pxietiming}{NI \pxie-6674T}
\newcommand{\pxi}{\mbox{PXI}}
\newcommand{\fpga}{\mbox{FPGA}}
\newcommand{\flex}{\mbox{FlexRIO}}
\newcommand{\labview}{\mbox{LabVIEW}}
\newcommand{\odaq}{\mbox{ODAQ}}
\newcommand{\odb}{\mbox{ODB}}
\newcommand{\caenhv}{\mbox{CAEN A1536}}
\newcommand{\caenmainframe}{\mbox{CAEN SY4527}}
\newcommand{\trgID}{\mbox{Trigger-ID}}
\newcommand{\trgIDs}{\mbox{Trigger-IDs}}

%% file: authors.tex
\author{
	P.~Agnes$^a$,
	L.~Agostino$^b$,
	I.~F.~M.~Albuquerque$^{c,d}$,
	T.~Alexander$^{i}$,
	A.~K.~Alton$^g$,
	K.~Arisaka$^h$,
	H.~O.~Back$^{c,i}$,
	B.~Baldin$^f$,
	K.~Biery$^f$,
	G.~Bonfini$^j$
% ...
	\\
	(The DarkSide Collaboration)
	\\
\llap{$^a$}APC, Universit\'e Paris Diderot, CNRS/IN2P3, CEA/Irfu, Obs de Paris, Sorbonne Paris Cit\'e, 75205 Paris, France\\
\llap{$^b$}LPNHE Paris, Universit\'e Pierre et Marie Curie, Universit\'e Paris Diderot, CNRS/IN2P3, Paris 75252, France\\
\llap{$^c$}Department of Physics, Princeton University, Princeton, NJ 08544, USA\\
\llap{$^d$}Instituto de F\'isica, Universitade de S\~ao Paulo, S\~ao Paulo 05508-090, Brazil\\
\llap{$^e$}Amherst Center for Fundamental Interactions and Department of Physics, University of Massachusetts, Amherst, MA 01003, USA\\
\llap{$^f$}Fermi National Accelerator Laboratory, Batavia, IL 60510, USA\\
\llap{$^g$}Department of Physics, Augustana University, Sioux Falls, SD 57197, USA\\
\llap{$^h$}Department of Physics and Astronomy, University of California, Los Angeles, CA 90095, USA\\
\llap{$^i$}Pacific Northwest National Laboratory, Richland, WA 99354, USA\\
\llap{$^j$}Laboratori Nazionali del Gran Sasso, Assergi AQ 67010, Italy\\
\llap{$^k$}Gran Sasso Science Institute, L'Aquila 67100, Italy\\
\llap{$^l$}Department of Physics, Universit\`a degli Studi, Genova 16146, Italy\\
\llap{$^m$}Istituto Nazionale di Fisica Nucleare, Sezione di Genova, Genova 16146, Italy\\
\llap{$^n$}Istituto Nazionale di Fisica Nucleare, Sezione di Milano, Milano 20133, Italy\\
\llap{$^o$}Istituto Nazionale di Fisica Nucleare, Sezione di Roma Tre, Roma 00146, Italy\\
\llap{$^p$}Department of Physics and Mathematics, Universit\`a degli Studi Roma Tre, Roma 00146, Italy\\
\llap{$^q$}Department of Physics, Universit\`a degli Studi, Cagliari 09042, Italy\\
\llap{$^r$}Istituto Nazionale di Fisica Nucleare, Sezione di Cagliari, Cagliari 09042, Italy\\
\llap{$^s$}Department of Physics, University of Houston, Houston, TX 77204, USA\\
\llap{$^t$}Istituto Nazionale di Fisica Nucleare, Sezione di Napoli, Napoli 80126, Italy\\
\llap{$^u$}Department of Physics, Universit\`a degli Studi Federico II, Napoli 80126, Italy\\
\llap{$^v$}Department of Physics, Virginia Tech, Blacksburg, VA 24061, USA\\
\llap{$^w$}Skobeltsyn Institute of Nuclear Physics, Lomonosov Moscow State University, Moscow 119991, Russia\\
\llap{$^x$}Department of Physics, Universit\`a degli Studi, Milano 20133, Italy\\
\llap{$^y$}St. Petersburg Nuclear Physics Institute NRC Kurchatov Institute, Gatchina 188350, Russia\\
\llap{$^z$}Department of Physics and Astronomy, University of Hawai'i, Honolulu, HI 96822, HI\\
\llap{$^{aa}$}Joint Institute for Nuclear Research, Dubna 141980, Russia\\
\llap{$^{bb}$}Department of Physics, Temple University, Philadelphia, PA 19122, USA\\
\llap{$^{cc}$}Kavli Institute, Enrico Fermi Institute and Dept. of Physics, University of Chicago, Chicago, IL 60637, USA\\
\llap{$^{dd}$}Institute for High Energy Physics, Beijing 100049, China\\
\llap{$^{ee}$}Laboratorio Subterr\'aneo de Canfranc, Canfranc Estaci\'on E-22880, Spain\\
\llap{$^{ff}$}IPHC,19 Universit\'e de Strasbourg, CNRS/IN2P3, Strasbourg 67037, France\\
\llap{$^{gg}$}School of Natural Sciences, Black Hills State University, Spearfish, SD 57799, USA\\
\llap{$^{hh}$}Institute for National Research, National Academy of Sciences of Ukraine, Kiev 03680, Ukraine\\
\llap{$^{ii}$}Radiation Physics Laboratory, Belgorod National Research University, Belgorod 308007, Russia\\
\llap{$^{jj}$}SLAC National Accelerator Laboratory, Menlo Park, CA 94025, USA\\
\llap{$^{kk}$}National Research Centre Kurchatov Institute, Moscow 123182, Russia\\
\llap{$^{ll}$}National Research Nuclear University MEPhI, Moscow 115409, Russia\\
\llap{$^{mm}$}Department of Chemistry, Biology and Biotechnology, Universit\`a degli Studi, Perugia 06123, Italy\\
\llap{$^{nn}$}Istituto Nazionale di Fisica Nucleare, Sezione di Perugia, Perugia 06123, Italy\\
\llap{$^{oo}$}Department of Physics, University of California, Davis, CA 95616, USA\\
\llap{$^{pp}$}Smoluchowski Institute of Physics, Jagiellonian University, Krakow 30059, Poland\\
\llap{$^{qq}$}Lawrence Livermore National Laboratory, Livermore, CA 94550, USA\\
\llap{$^{rr}$}Institute of Physics Gleb Wataghin  Universidade Estadual de Campinas, S\~ao Paulo 13083-859, Brazil\\
\llap{$^{ss}$}School of Physics, University of Chinese Academy of Sciences, Beijing 100049, China\\
}

%% file: ds50experiment.tex
\section{The DarkSide experiment}\label{sec:ds50experiment}

Many experimental results in cosmology and astrophysics provide evidence for the existence of a gravitationally-interacting, non-luminous dark matter.
One of the favored candidates for dark matter are weakly interacting massive particles (\wimps), neutral particles with mass $\sim 100$\,GeV and cross-section for interaction with nucleons in the range $\sim 10^{-45}\,\mbox{cm}^2$ to $\sim 10^{-48}\,\mbox{cm}^2$ that can be gravitationally bound inside our galaxy~\cite{goodman}.
If \wimps\ exist, they should occasionally interact with an atomic nucleus, causing the nucleus to recoil with kinetic energy less than $100$\,keV.

The \dsf\ experiment attempts to detect \wimp-induced nuclear recoils using a two-phase liquid argon time projection chamber (\lar-\tpc), operated at the Gran Sasso National Laboratory (LNGS) in Italy.
A WIMP search with an exposure of \dsfexpoaar\ using atmospheric argon (\aar) has been performed~\cite{ds50first}.
\dsf\ is now conducting a dark matter search with the \lar-\tpc\ filled with underground argon (\uar), which contains a substantially reduced content of radioactive \ar.
The first WIMP search using \uar\ with an exposure of \dsfexpouar\ is presented in~\cite{ds50second}.

A key feature of the \dsf\ design is its active veto system, composed of a liquid scintillator veto (\lsv), serving as shielding and as anti-coincidence for radiogenic and cosmogenic neutrons, \grs, and cosmogenic muons, and a water Cherenkov veto (\wcd), acting as  passive shielding and as anti-coincidence for cosmogenic muons~\cite{bxmuonflux, bxcosmogenic}.
Detailed information on the veto detectors of \dsf\ can be found in~\cite{ds50veto}.
Simulation results about the effectiveness of this veto systems are reported  in~\cite{veto1, ae_evh} and in the included references.
This paper describes the electronics and data acquisition system (\daq) of the active veto system of the \dsf\ experiment.

%% file: ds50vetoes.tex
\section{The \dsf\ veto detectors} \label{sec:ds50vetos}

\begin{figure}[tp]
\begin{center}
\includegraphics[width=0.5\columnwidth]{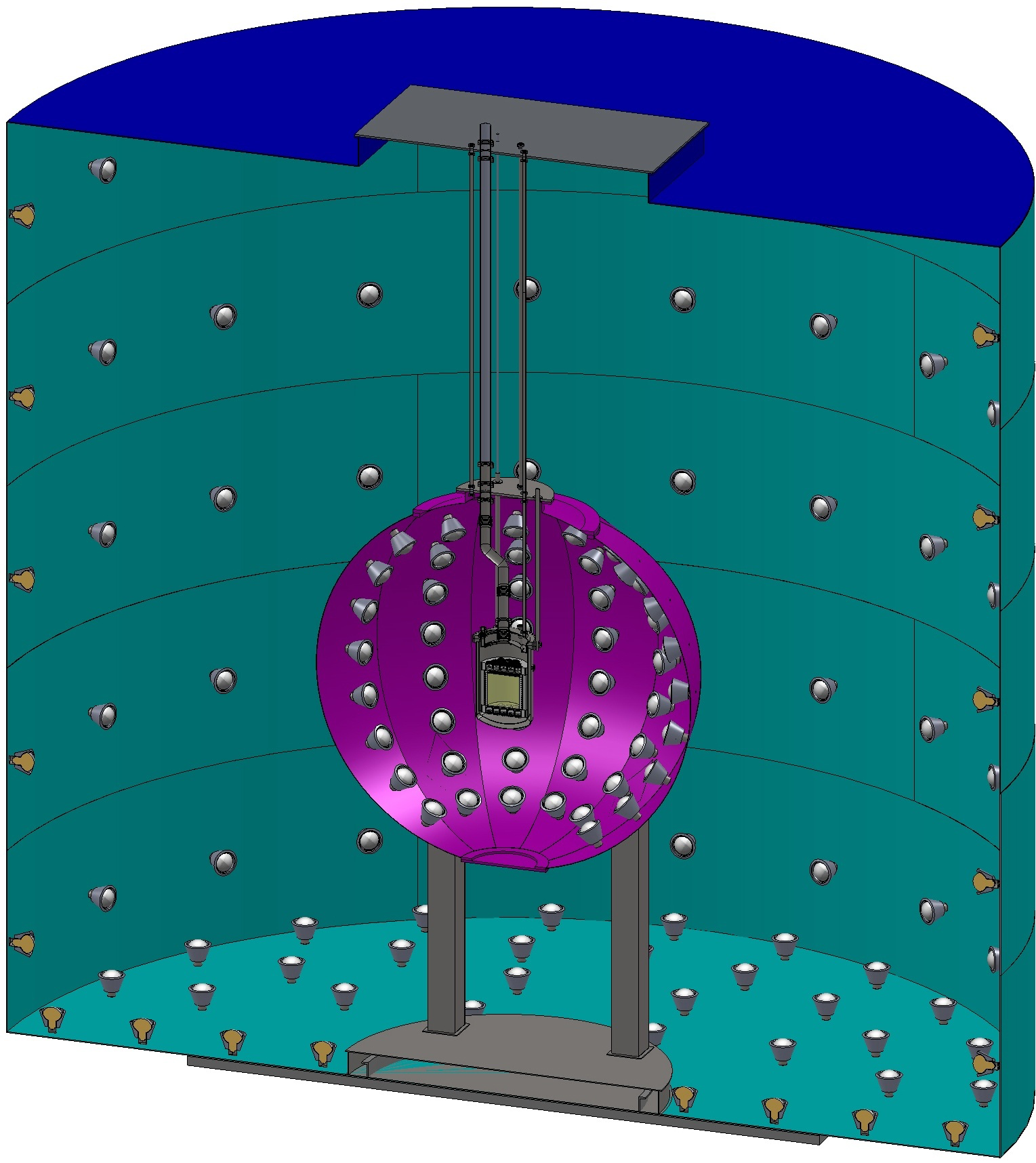}
\caption{The \dsf\ experimental layout showing the mechanical scheme of the \wcd, the \lsv, and the \lar \tpc.}
\label{dsexp}
\end{center}
\end{figure}

The \dsf\ apparatus consists of three nested detectors (see figure~\ref{dsexp}). 
From the center outward, the three detectors are: the liquid argon \tpc, which is described in detail in~\cite{ds50first, ds50second}, the \lsv\ and the \wcd.
The \lsv\ and \wcd\ detectors constitute the active veto system of \dsf~\cite{ds50veto}.
The \dsf\ detector system is located in the Hall C of LNGS at a depth of $3800$\,m.w.e., in close proximity and sharing many facilities with the solar neutrino detector \bx~\cite{bxlong, bxdetpa, bxfluid}.
In fact, the tank  of the \wcd\ is the same tank that was used for the prototype of the \bx\ experiment~\cite{ctf}.

The electronics room for the \dsf\ veto detectors is located next to the electronics room of the \bx\ experiment.
Readout of the veto signals from the \lsv\ and \wcd\ are patched through the side of the \wcd\ tank to the veto electronics room by cables $\sim40$\,m in length,
as described in detail in~\cite{ds50veto}.

\subsection{The \lsv} \label{sec:lsv}

The \lsv\ detector is a $4$\,m diameter stainless steel sphere filled with $30$\,tonnes of boron-loaded liquid scintillator.
The scintillator is a mixture of pseudocumene (\pc) and trimethyl borate (\tmb, \tmbchem), with $2,5$ diphenyloxazole (PPO) as fluor.
The interior surface of the \lsv\ sphere is covered with Lumirror diffusive reflecting sheets~\cite{lumirror}. 
The \lsv\ is instrumented with an array of \lsvpmtnum\ \lsvpmt\ \lsvpmtsize\ \pmts~\cite{ham5912}, with low radioactivity glass bulbs and high-quantum-efficiency photocatodes (\lsvpmtqe\ average quantum efficiency (QE) at \lsvpmtwave).
The phototube coverage of the \lsv\ is $\sim7\%$.
The detailed description and characterisation of the \lsv\ \pmts\ can be found in~\cite{ds50veto}.

The \lsv\ is designed to identify and veto neutrons which might enter or exit the \lar\ \tpc.
Neutrons thermalize by scattering on protons in the liquid scintillator and are efficiently captured by \borten\ nuclei via two channels:
\begin{center}
\begin{tabular}{llc}
$^{10}$B + n $\rightarrow$ $^7$Li + $\alpha$ & & 6.4\% \\
$^{10}$B + n $\rightarrow$ $^7$Li$^{*}$ + $\alpha$, & $^7$Li$^{*}$ $\rightarrow$ $^7$Li + $\gamma$ (478\,keV) & 93.6\% \\
\end{tabular}
\end{center}
Capture on \borten\ proceeds to the \lith\ ground state with a branching ratios of 6.4\%, producing a $1775$\,keV $\alpha$ particle, and to a \lith\ excited state with branching ratio 93.6\%, producing a $1471$\,keV $\alpha$ particle and a $478$\,keV \gr.
Neutrons can also capture on hydrogen, which causes the emission of a $2.2$\,MeV \gr.

The \tmb\ contains natural boron with a \abborten\ natural abundance of \borten\, which has a thermal neutron capture cross section of 
\lsvbtenxsec~\cite{veto1}.
Loading \tmb\ in the \pc\ thus shortens the thermal neutron capture time.
The thermal neutron capture time in a pure \pc\ scintillator is $\sim250$\,$\mu$s; it becomes  $\sim2$\,$\mu$s ($\sim22$\,$\mu$s) in
a 50\% (5\%) mixture of \pc\ and \tmb, as in the first (second) phase of \dsf~\cite{ds50first, ds50second, ds50veto}.

Ionizing events in the \lsv\ produce scintillation photons. These photons propagate inside the detector undergoing diffusive reflections on the Lumirror and the \tpc\ cryostat until they are eventually detected by the \pmts.  
Due to the multiple reflections and low phototube coverage, the time distribution of the light signal extends up to $300$\,ns.

The information about the energy deposited in the \lsv\ by a scintillation event is contained in the total charge collected by all \pmts\ within this time interval.
The measured light yield in the \lsv\ is $\sim0.5$\,photoelectrons/keV (PE/keV) for $\beta$ and \grs\ in the energy range from a few tens of keV up to a few MeV.
The scintillation light produced by $\alpha$ particles is heavily reduced by scintillation quenching. 
In particular, the light output of the $1775$\,keV $\alpha$ from the neutron capture on \borten\ has a $\beta$-equivalent energy of $50$-$60$\,keV, corresponding to $20$-$30$\,PE detected by the \pmts\, as measured in~\cite{ds50veto}. 

The \pmts\ work in the single photoelectron regime for scintillation events depositing less than $\sim200$\,keV in the \lsv, although the probability of having more than one photoelectron in a given \pmt\ is not completely negligible, even for relatively low energy deposits.
The signal due to a single photoelectron at the \pmt\ output is a pulse with an amplitude of about $12$\,mV (on a \mbox{50}\,{$\omega$} load) and $\sim$20\,ns total width.
The \pmts\ are AC coupled and grounded at the cathode, so the meaningful part of the signal has negative polarity.

Scintillation events can happen anywhere in the volume of the \lsv.
Scintillation happening far from \pmts\ will deposit approximately the same fraction of light on each \pmt\ in the detector, due to the multiple reflection of the light.
However, this fraction can increase drastically for events happening near or direclty in front of a \pmt.
Additionally, when a muon crosses the \lsv, a huge scintillation signal is produced, corresponding to an energy deposit of $\sim2$\,MeV per cm of scintillator traversed. 

\subsection{The \wcd} \label{sec:wcd}

The \wcd\ is a cylindrical stainless steel tank, 11\,m in diameter and 10\,m high, filled with $1000$\,tonnes of ultra-pure water. 
To maximize the number of photons collected by the \pmts\, the internal surface of the tank is covered with reflecting Tyvek sheets.
The \wcd\ is instrumented along the floor and on the side of the cylindrical wall with an array \ctfpmtnum\ \ctfpmt\ \ctfpmtsize\ \pmts~\cite{ctf}, with \ctfpmtqe\ average QE at \ctfpmtwave. More information on the \wcd\ can be found in~\cite{ds50veto}.

The \wcd\ serves two functions: it is a passive shield against external \grs\ and neutrons, and it is an active Cherenkov detector for muons crossing the \lar\ \tpc\ or passing close enough to produce dangerous background events through the spallation of of the various nuclei in the detectors.

When a muon crosses the \wcd, a huge Cherenkov signal is produced. 
The signal is usually evenly distributed among all the \pmts\ because of the multiple reflections on the Tyvek and the low photocathode coverage.
Due to the magnitude of the muon signal (larger than about 400\,PE) and  fast rise time (about 20\,ns), muon events are easily distinguishable from noise events.
For the \wimp\ dark matter search, we are interested in rejecting \lar\ \tpc\ events that coincide with muon-like events in the \wcd.

%% file: vetorequirements.tex
\section{Requirements for the veto electronics and \daq}
\label{sec:vetorequirements}

Following  very general considerations in the section above, we have designed the \lsv\ and \wcd\ electronics and \daq\ according to the following requirements.

\begin{itemize}
	\item \emph{Single photoelectron detection with high efficiency and good time resolution}: the front-end (FE) and the digital electronics must be able to efficiently detect a signal as small as $0.25$\,PE with a time resolution  better than $1$\,ns. 
This sub-PE threshold allows the detector to achieve high detection efficiency for very low energy events, including in particular the neutron capture signals on \borten, or short tracks in the \wcd. 
The good timing can be useful for eventual position and track reconstructions through time of flight techniques, following what has already been done, for example, by the \bx\ experiment~\cite{bxlong, bxmuon};
	\item \emph{Large dynamic range}: although most of the interesting events occur at low energy, it is important that the system performs well even when a huge signal appears in the scintillator, like the one generated by a muon or a muon-induced shower. 
It is difficult (and unnecessary in this case) to avoid saturating either  the analogue amplifier of the ADCs, but it is very important that the channel has a relatively short recovery time, so that muon-induced events can be studied efficiently. 
We require that the vertical dynamics of the system can be as large as $7$\,PE and that the system be able to recover from a very large ($\times$1000 or more) signal in less that $20$\,$\mu$s;
	\item \emph{Synchronisation among channels}: it is important that the relative timing among hits collected by different \pmts\ (either in the \lsv\ or in the \wcd) is known and stable over the whole duration of a run, which typically continues for several hours. 
We require that the channels are synchronized within 1\,ns, and that this synchronization is stable for at least 24 hours of continuos data taking;  
	\item \emph{Triggering capability and \tpc\ synchronization}: the veto detectors are operated during \tpc\ runs. 
Two independent modes should be possible: one in which the veto is triggered by the \lar\ \tpc\ (regardless of the veto activity) and one in which the veto triggers on the \lsv\ or \wcd\ activity (regardless of the \lar\ \tpc\ activity).
\tpc\ and veto events should be synchronized within at least $20$\,ns.
The veto system should be able to trigger independently and the dead-time between two consecutive triggers should be virtually zero to avoid inefficiencies, particularly for correlated events\footnote{This paper describes the current implementation of these features. A project is in progress to implement trigger-less data taking of all \pmt\ pulses with zero dead time and implementation of a pure software trigger. This new feature will be described in a separate paper.}.
\end{itemize}

\begin{figure}[tb]
\begin{center}
\includegraphics[width=0.6\columnwidth]{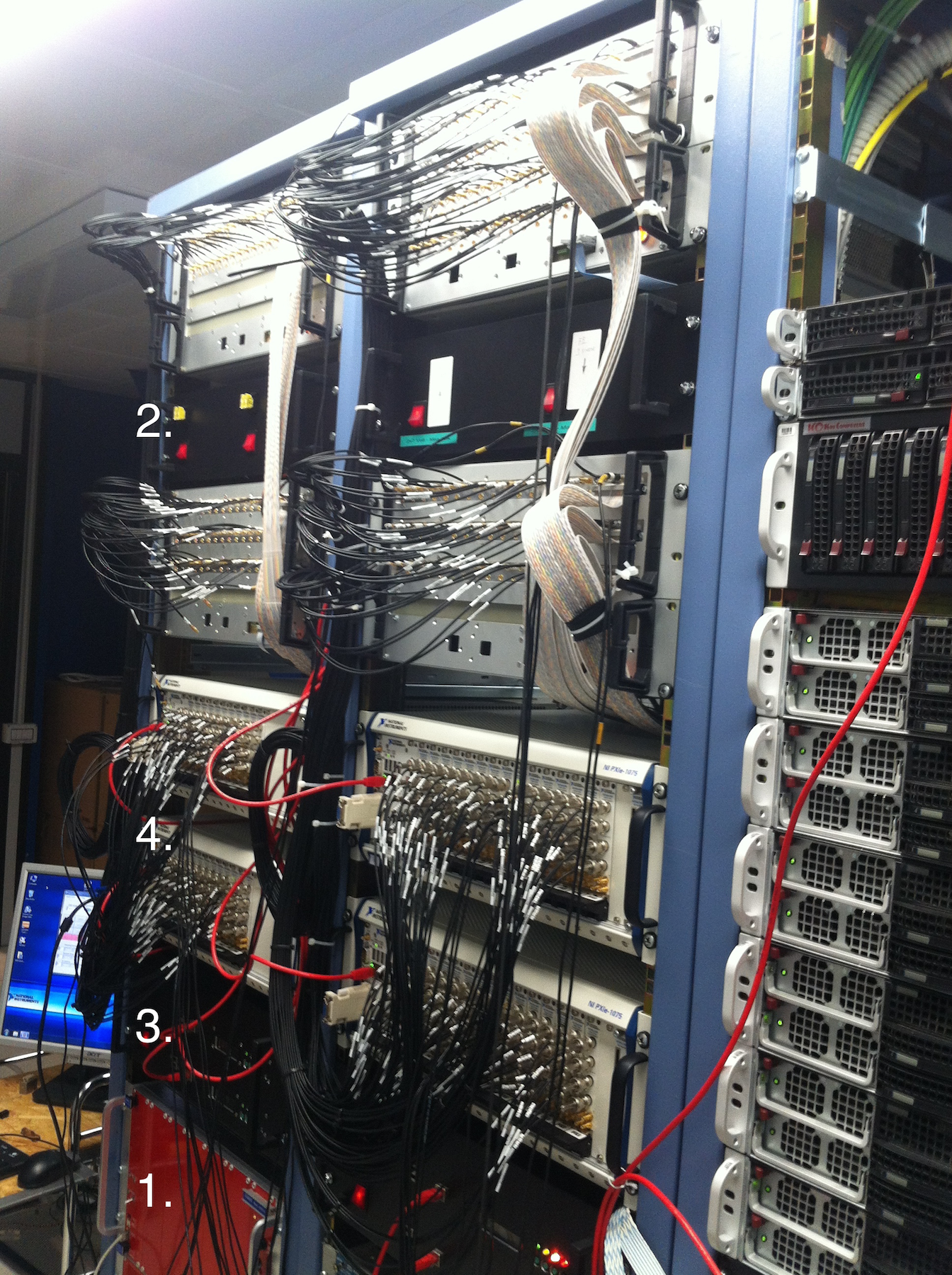}
\caption{A picture showing the \dsf\ veto electronics in place.
1: the CAEN Mainframe (\caenmainframe) for the CAEN HV boards (\caenhv); 2: the custom built front-end modules (FEM); 3: the custom built front-end digital board (FEDB); 4: the commercial National Instruments chassis (\pxiechassis) housing the National Instruments digitizers (\pxiedigitizer).}
\label{fig:vetoele_pic}
\end{center}
\end{figure}

We have implemented these requirements in a system made of a custom built FE electronics and commercial National Instruments (NI) high speed digitizers, shown in figure~\ref{fig:vetoele_pic}. 
The same front-end modules (FEM) and digitizers are used for both channels of the both \lsv\ and \wcd. 

%% file: vetofe.tex
\section{Veto front-end electronics} \label{sec:fe}

\begin{figure}[tbp]
\begin{center}
\includegraphics[width=0.6\columnwidth]{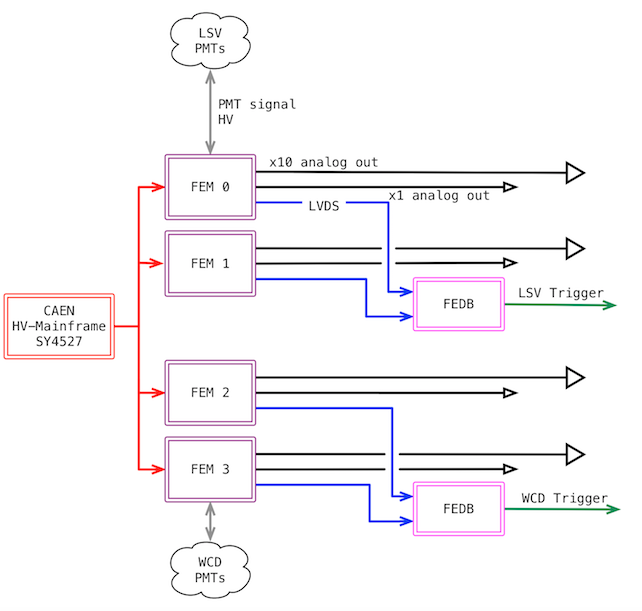}
\caption{A block diagram of the front-end system of the \dsf\ veto.}
\label{fig:fe_scheme}
\end{center}
\end{figure}

\begin{figure}[tbp]
\begin{center}
\includegraphics[width=0.9\columnwidth]{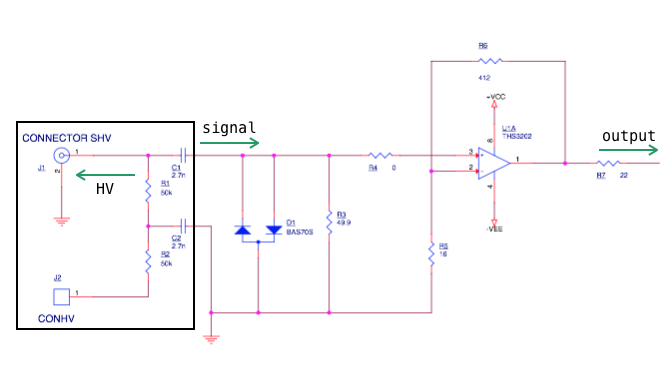}
\caption{Circuit diagram of AC coupling and amplification in the \feab.}
\label{fig:fe_circuit_zoom}
\end{center}
\end{figure}

\begin{figure}[tbp]
\begin{center}
\includegraphics[width=0.95\columnwidth]{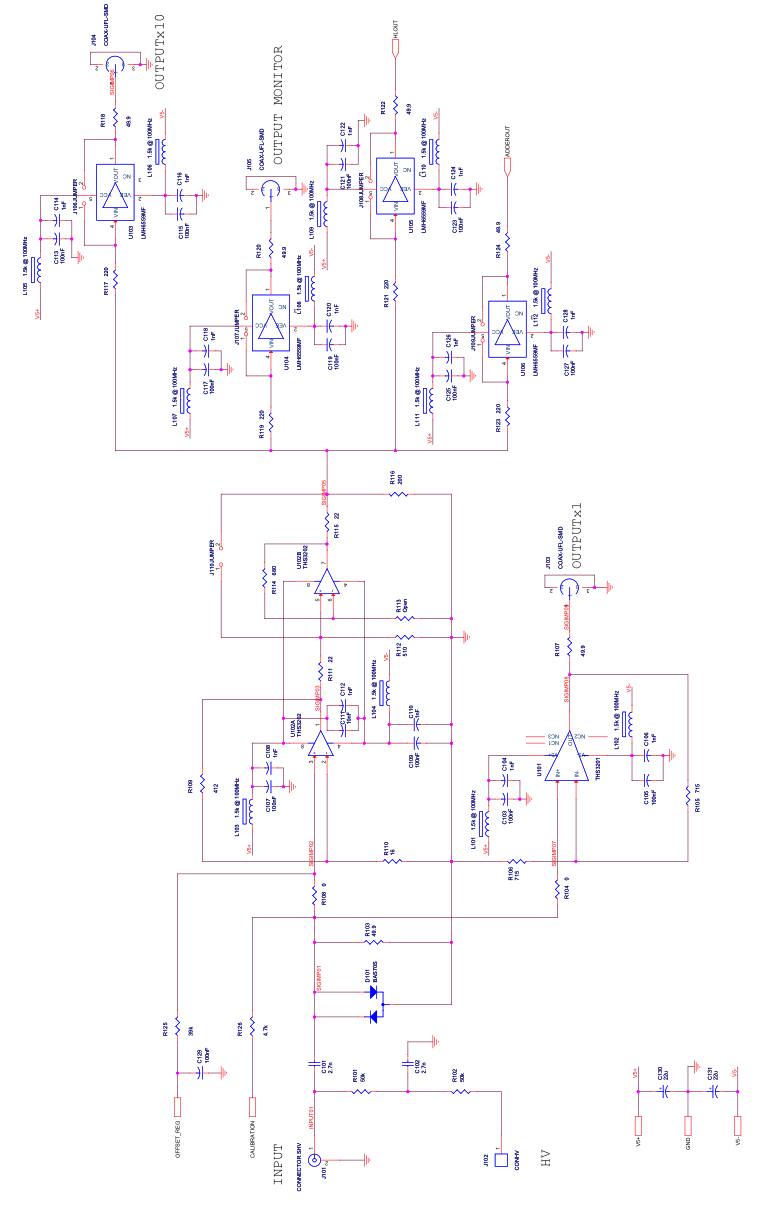}
\caption{The full circuit diagram of the \feab.}
\label{fig:fe_circuit}
\end{center}
\end{figure}

The FE electronics for the \dsf\ \lsv\ and \wcd\ realize a set of functions that can be divided into analog and digital functions.

The FE analog functions are:
\begin{itemize}
\item due to the fact that \lsv\ and \wcd\ \pmts\ are AC coupled, FE must provide the high voltage and decouple the signals along a single cable;
\item the PMT signals must be amplified by about a factor of 10 to achieve enough resolution on the single PE response (SER). 
The amplification must be at high bandwidth in order not to spoil the fast rise time of the SER ($\sim3$\,ns);
\item in order to have consistent dynamic range of the digitizers across the channels, the FE must provide programmable input offset compensation;
\item a second non amplified output is very useful to extend the energy range of the detector;
\item easy scope inspection of the \pmts\ must be allowed through a second front panel output;
\item a sum of the output of 16 channels  is needed to build all channel sums for additional acquisition or analog trigger purposes;
\item the possibility to distribute signals from a calibration input to all channels must be ensured.
\end{itemize}

The FE digital functions are:
\begin{itemize}
\item the signals must be discriminated on-board with programmable threshold to provide the digital signals used for the trigger and the monitor of the single PMT count rate. 
The discriminated signals are also fed to long range TDCs within the \tpc\ acquisition process (see section~\ref{sec:tdc}).
\item a monitor of the single \pmt\ count rate must be provided;
\item the \lsv\ and the \wcd\ trigger signals must be generated whenever a programmable number of channels in each sub-detector is firing within a time window of programmable duration;
\item the run controller must be able to connect via TCP/IP to the FE in order to set parameters and to read single \pmt\ count rate data;
\item a local human interface with a display, buttons and LEDs is needed for prompt feedback on the system.
\end{itemize}

In order to fulfil these requirements we have designed a system composed of front end analogue boards (\feab) and front end digital boards (\fedb).
The system is composed of two identical modules for each sub-detector. 
Each 5U 19'' module hosts up to four \feabs\ (\emph{i.e.} up to 64 channels) and is mastered by a \fedb\ and features a dedicated linear power supply.
A block diagram of the FE system is in figure~\ref{fig:fe_scheme}.

\subsection{Veto \feab\ design}
\label{sec:vetofeabdesign}

The \feabs\ were designed and implemented to provide the functionality expressed above with a high density of channels and low added noise.
After simulations performed with PSPICE~\cite{PSPICE} and tests on a prototype, we came to the current configuration which houses 16 channels on each board, each with one input and three outputs.
The final design has $200$\,$\mu$V$_{\mbox{rms}}$ total output noise in its $230$\,MHz bandwidth at $20$\,dB gain setting.

A circuit diagram of the HV decoupling and amplification stage of the \feab\ is shown in figure~\ref{fig:fe_circuit_zoom}, while figure~\ref{fig:fe_circuit} shows the full \feab\ circuit diagram.
The two resistors R1 and R2 in figure~\ref{fig:fe_circuit_zoom} supply the high voltage to the PMT.
The capacitor C1 ($2.7$\,nF) acts as a high-pass filter, blocking the DC high voltage from the components of the board downstream.
Due to this AC coupling, the downstream signal becomes bipolar.
The time constant of the RC coupling is $135$\,ns.

The THS3201 (single) and THS3202 (double) fast current feedback operational amplifiers~\cite{TI3201} are mounted on the FEAB, whose gain bandwidth product is $1.8$\,GHz.
The signal is split into two branches, using the THS3202 for two gain $\times$10 paths and the THS3201 for a gain $\times$1 path. 
High-speed LMH6559 buffers~\cite{LMH6559} provide fan-outs to the $\times$10 output (given to the digitizers), the $\times$10 monitor output (oscilloscope probe), the discriminator, the offset regulation, and the adder output (sum of 16 channels).
To host 16 channels in 1 PCB layout, we arranged 2 rows of 8 channels.
\feab\ features 8 PCB Layers, with a $274 \times 274$\,mm footprint containing 2169 components on the top and bottom of the board.

On-board discrimination is implemented using the ADCMP567~\cite{compar}, a dual ultrafast voltage comparator with differential PECL compatible output, followed by a buffer to send the digitized output (in LVDS standard) to the digital board. 
The threshold is set by the 8-bit DACs using the I$^2$C protocol.

The front panel top row alternates MCX and LEMO connectors for the $\times$10 outputs to be connected to digitisers and for scope inspections, respectively, while the bottom row features the MCX connectors for the $\times$1 outputs.
The last two LEMO connectors on the left side of the pannel are the calibration input and the sum output. 
The 34-pin connector yields  LVDS discriminated outputs to the \fedb.
The rear panel features 16 HV connectors where \pmt\ cables are directly connected and a REDEL KAG.H22 multi-pin connector to the HV boards \caenhv~\cite{caenhv} housed in a CAEN Mainframe SY4527~\cite{caenmf}.

Special care has been devoted to minimize the noise originating from the power supplies. 
We avoided the use of switching power supplies and  built a linear power supply able to deliver $15$\,A at the dual voltage, $\pm7.5$\,V, required for the \feab\ modules.

\subsection{Veto \fedb\ design}

The \fedb\ handles the digital functions expressed above.
It houses a Xilinx Spartan 6 \fpga~\cite{spar6} to perform the fast actions following the discriminator firing, namely   trigger formation and the measurement of the single \pmt\ rate. 
Slow control operations are implemented by a PIC32 micro-controller~\cite{micro_controller} mounted on its ethernet starter kit and  interconnected to the \fpga.
Both the \fpga\ firmware and the micro-controller program were developed specifically for \dsf.

The \fedb\ houses a $100$\,MHz local clock issued by a quartz oscillator. 
The Xilinx Spartan 6 \fpga\ can operate on the local clock, however synchronization cannot be guaranteed in this way. 
For this reason, the \fedb\ hosts Micro Miniature MQ172 input connectors  receiving a distributed clock signal.

The \fpga\ receives the 64 TTL logic inputs originated from the discriminators housed on the \feabs. 
The LVDS output of the \feab\ are brought to a piggy-back board of the \fedb\ where they are duplicated to feed the TDC and then converted to single-ended TTL standard for the \fpga.
At each clock cycle, the \fpga\ saves the status of the TTL lines and compares it with the previous one. 

\subsubsection{\fedb\ single \pmt\ rate function} 
\label{sec:feab_scaler}

At each clock cycle, if a transition from state 0 to state 1 is detected on a TTL input line, a 16-bit counter corresponding to that channel is incremented.
Every $100$\,ms the 64 counters are copied to a cache and then reset. 
Whenever the user requests the counts for a channel the cached copy is provided.
If a channel exceeds a programmable threshold $N_{hot}$, it is declared `hot' (noisy). 
$N_{hot}$ can be set in the $10-2550$\,kHz range.
A front panel orange or red LED is lit if there is at least one or ten hot channels, respectively.
If the user enables the specific functionality, the hot channels can also temporarily be excluded from the trigger while in the `hot' condition.

\subsubsection{\fedb\ trigger function} 
\label{sec:fedb_trg}

In both the \lsv\ and the \wcd, \fedbs\ are daisy chained in a master slave configuration for triggering purpose, so that the generation of a trigger in the master is done on the full sub-detector rather than on just the 64 channels physically handled in the crate, while the trigger condition, evaluated in the slave, is ignored. 
Although two \fedb s are enough for \dsf, the system is designed to scale up to larger detectors.
In fact, each crate masters up to 4 slaves and more then two levels of cascading can be used.

In each \fpga, at each clock cycle, if a transition from state 0 to state 1 is detected on a TTL input line, a numeric flag for that channel is set to a programmable number $N_{win}$, unless the channel is disabled by the user (command set flag) or due to a high rate (see above).
All flags different from zero are decremented at each clock transition. 
Each channel contributes to trigger formation for $N_{win}$ clock cycles.
At each transition, the number of flags different from zero is computed, stored in a 16 events FIFO and provided as a 10-bit sum output on an external connector along with a copy of the clock signal.
The sum output is fed to the summed input present in the following module.
Each \fpga\ samples the sum inputs (four are present for scalability) on the incoming clocks and adds them. 
The resulting record is added to sum of the internal channels retrieved from the FIFO with a programmable depth $N_{del}$ ($N_{del}$=5 in our setup if running at $100$\,MHz). 
This mechanism ensures time alignment between the two crate stages by taking into account the cable length and the clock cycles needed in the master crate in order to sum the input signals from the preceding slave crates.
Finally, the sum of all channels is compared to a programmable threshold $N_{thr}$ eventually determining a trigger condition.

An internal prescale mechanism is also implemented to handle possible high trigger rates at low threshold.
Upon meeting the trigger condition a down counter is decremented. A trigger is issued only when the counter reaches 0, after which it is reset. 
The counter value $N_{pre}$ is programmable up 1024. 
The default value of 1 determines no prescale, with triggers being issued every time the condition is met.

The duration of the trigger length is also handled via a programable down counter $N_{len}$ in the $100$\,ns-$25.5$\,$\mu$s range. 
While the trigger signal is on, the trigger condition is not checked, resulting in a self-vetoing behaviour. 
A yellow LED on the front panel is lit for $40$\,ms whenever a trigger is issued. 
The number of triggers is recorded with a dedicated counter, copied to a cache counter and reset every $100$\,ms.
The content of the cached counter can be returned to the micro-controller upon request.

\subsubsection{FE slow control functions} 
\label{sec:fe_slow}

The \fpga\ is interfaced to the micro-controller via numerous I/O lines. 
A set/get command protocol is implemented with 16-bit data out (\fpga\ to micro-controller and 10-bit data in (micro-controller to \fpga), plus a few control lines. 
The bus is mastered by the micro-controller and data are always polled. 
Commands are numerically coded in mirror tables between micro-controller and \fpga\ codes.

The micro-controller runs a single process which acquires an IP address from the LAN DCHP and acts as command interpreter by waiting for incoming socket connections on a specific port.
In additions to the commands resolved by polling the \fpga, the micro-controller also communicates via I$^2$C to DACs for setting and reading back the channel input offset compensation (12-bit) and the discriminator thresholds (8-bit). 
The micro-controller also controls via I$^2$C an alphanumeric display mounted on the front panel (2 lines of 16 characters each). 
This is used to display single channel information (rate, threshold, offset, contribute-to-trigger-status) or trigger information (primary screen: rate, level, window length, prescale, secondary screen: trigger length, number of channels contributing, delay, status of automatic disabling). 
The command interpreter is interrupted by the timer every $500$\,ms to reread all displayed parameters and to refresh the screen. 
A green LED is toggled during this interrupt, obtaining a blinking ``alive'' monitor.  
Four buttons are read by interrupt request on the micro-controller which allow it to switch between single channel and trigger screens and to change the displayed channel.

%% file: vetodaq.tex
\section{Veto \daq} 
\label{sec:veto_daq}

\begin{figure}[tbp]
\begin{center}
\includegraphics[width=0.98\columnwidth]{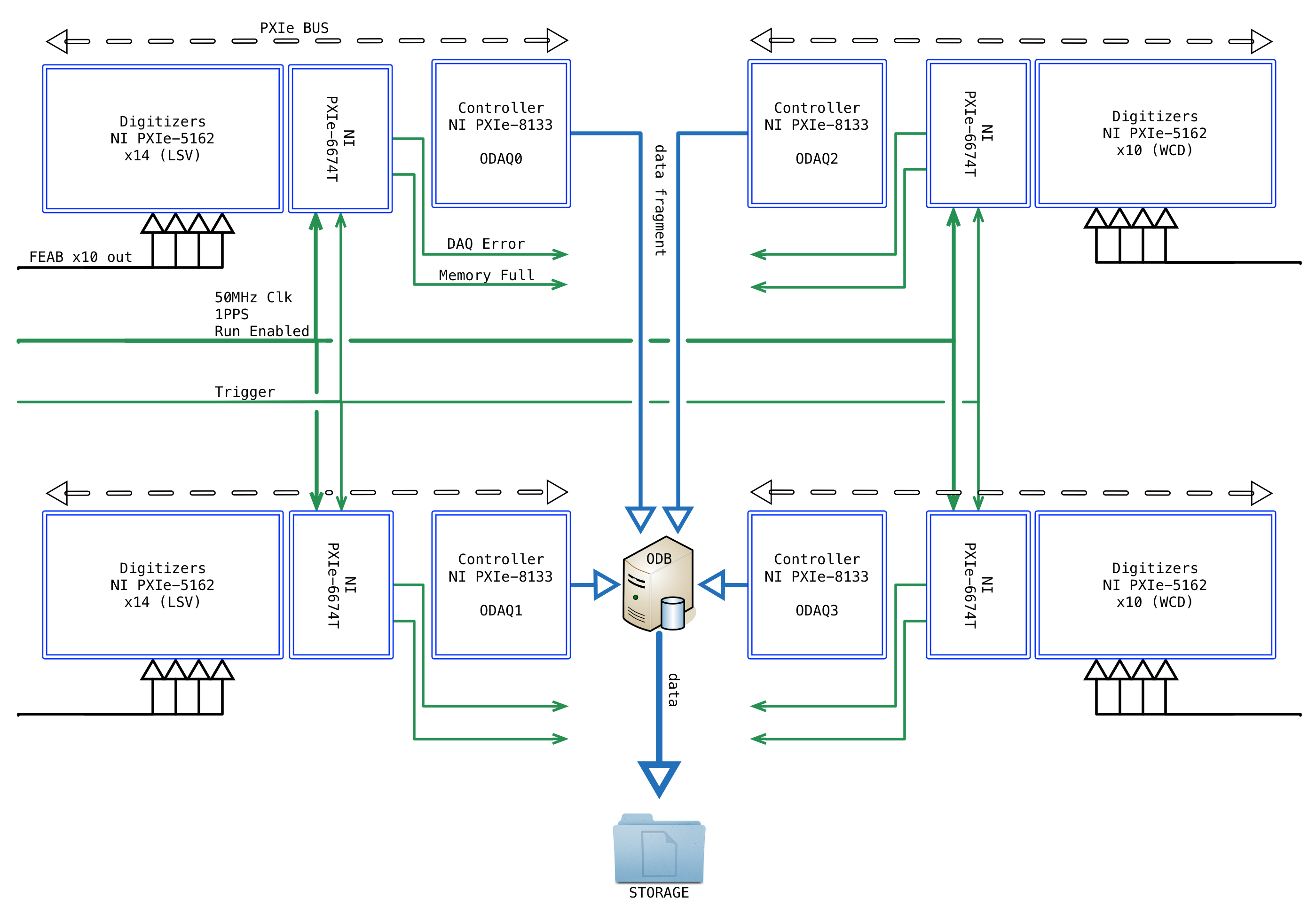}
\caption{A block diagram of the \daq\ system of the \dsf\ veto.}
\label{fig:veto_daq}
\end{center}
\end{figure}

A block diagram of the \daq\ for the veto system of \dsf\ is displayed in figure~\ref{fig:veto_daq}.
The analogue signals of all 190 \pmts\ in the veto are digitized and acquired using commercial NI \pxie\ digitizers after the amplification in the front-end stage.
Waveform samples are acquired by the \daq\ only when a trigger is received.
The data readout code (\odaq, outer detector \daq) runs on each \pxie\ controller.
When a trigger is received, \odaq\ reads the waveforms from the digitizers and performs zero suppression.
Four \pxie\ controllers are used, one for each \pxie\ chassis, in order to handle the 190 channels of the veto system.
Each instance of \odaq\ transmits the data fragment to the event builder PC via TCP/IP.
The event builder application (\odb, outer detector builder) collects each data fragment from the four \pxie\ controllers and writes the event to disk.

The following sections describe the main features of the veto \daq\ hardware, architecture and trigger.

\subsection{Veto \daq\ design} 
\label{sec:veto_daq_design}

The veto \daq\ software system consists of two distinct applications: the acquisition-and-readout application \odaq\ and the event builder application \odb.
An instance of \odaq\ runs on each of the 4 \pxie\ controllers. 
\odb\ runs on a devoted PC. 

The \daq\ of the \dsf\ veto has been designed to fulfil the set of requirements listed below:
\begin{itemize}
\item the system acquires data from 4 different \pxie\ chassis, preserving synchronization between each device in use;
\item the \daq\ system should be scalable (in principle) to any number of \pxie\ chassis;
\item all \daq\ applications are based on a state machine which correctly performs the necessary operations ({\it e.g.} initialize hardware, fetch data, stop acquisition, etc.);
\item all \daq\ applications should act as a server to listen for commands from the run controller (described in section~\ref{sec:rc}), such as a request to start and stop data acquisition, communicating the system status, and provide the number of acquired events;
\item the veto \daq\ system should sustain an input data throughput (prior to zero suppression) up to $1$\,GB per second on each \pxie\ chassis;
\item \odaq\ must perform zero-suppression on the acquired data;
\item \odaq\ must provide host-target communication with the \fpga\ module, to retrieve the \trgID\ and the time-stamp of the events;
\item all instances of \odaq\ running on the 4 \pxie\ controllers can transmit data over a LAN to \odb;
\item \odb\ must check the consistency of the time-stamps of the events, then bundle zero-suppressed data from different chassis in a data fragment structure, which contains the time-stamps and \trgID\ of the event and the zero-suppressed waveforms;
\item all \daq\ applications should check CPU and memory usage in the machines running the \daq, and gracefully make a transition to an error state if these exceed preset limits;
\item \odaq\ must periodically check the temperature of all digitizers and automatically shut them down to avoid potential thermal harm to the devices.
\end{itemize}

Both \odaq\ and \odb\ are written in \labview~\cite{labview} and conform to system design approach and documentation standards. We decided to use \labview\ for the following reasons. It provides a natural integration with NI hardware, the availability of essential libraries, modules, and toolkits to develop a daq system ({\it e.g.} TCP/IP libraries) with a customizable user interface, high level design tools for multithreading and FPGA. This choice resulted in a fast development time with a small team of developers.

\subsection{Veto \daq\ hardware} 
\label{sec:veto_daq_elect}

The veto \daq\ electronics provides digitized waveforms for each $\times$10 FE channels.
It is organised in 4 \pxie\ chassis: 2 for the \lsv\ and 2 for the \wcd.
Each chassis houses analog inputs for the waveform digitizers (56 for the \lsv, 40 for the \wcd) and digital TTL lines for the trigger and synchronization interface.

Each \pxie\ chassis (\pxiechassis\ chassis~\cite{NI1075}) is equipped with one \pxiecontroller\ controller~\cite{NI8133}, up to 14 digitizer boards \pxiedigitizer~\cite{NI5162}, one \pxietiming\ timing and synchronization module~\cite{NI6674} and one NI \pxie-7961R \flex\ \fpga\ module~\cite{NI7961}.

The \pxiecontroller\ controller is a high-performance Intel Core i7-820QM processor-based embedded controller for use in \pxi\ Express systems with $1.73$\,GHz base frequency, $3.06$\,GHz quad-core processor, and dual-channel $1333$\,MHz DDR3 memory.
The \pxie\ controller is a true computer with Windows 7 OS.
Each \pxie\ controller is equipped with two Gigabit Ethernet ports, used to transmit data to \odb\ and to receive commands from the run controller of the \dsf\ experiment.

The waveform digitizers  are \pxiedigitizer\ modules.
The sampling speed is $1.25$\,GSample per second (period $800$\,ps) and the resolution is 10-bit. 
Each sample is then 2 bytes in size.
Each waveform digitizer module has 4 BNC input channels.
The total on board memory is $1$\,GB ($256$\,MB per channel). 
The input vertical ranges for each channel can be selected as $0.1$, $0.2$, $0.5$, $1$, $2$ or $5$\,V peak to peak and
the vertical offset can be regulated within the vertical range.
The amplitude range is usually set between $+0.1$\,V and $-0.9$\,V, corresponding to a vertical range of $1$\,V with an offset of $-0.40$\,V.

Trigger and synchronization functions are performed with \pxietiming\ timing modules, with one in each \pxie\ chassis.
The timing module routes the clock and the trigger signals to the waveform digitizers and to the \fpga\ module in the \pxie\ chassis, allowing synchronization between all devices in the \pxie\ chassis. The timing module also routes the digital signals needed to generate timestamps to the \pxie-7961R \flex\ \fpga\ module. 
The I/O connectors of the timing module are female SMA. 
The digital signal logic is TTL.

One NI \pxie-7961R \flex\ \fpga\ module is used to generate the timestamp of each event using a $50$\,MHz reference clock common to the whole \dsf\ experiment, a 1 pulse per second (1PPS) signal received from the LNGS GPS, and a run enabled signal.
This module also handles two digital lines to communicate \daq\ errors and memory full signals.

\subsection{\daq\ software architecture and trigger} 
\label{sec:veto_daq_arch}

\begin{figure}[tbp]
\begin{center}
\includegraphics[width=0.85\columnwidth]{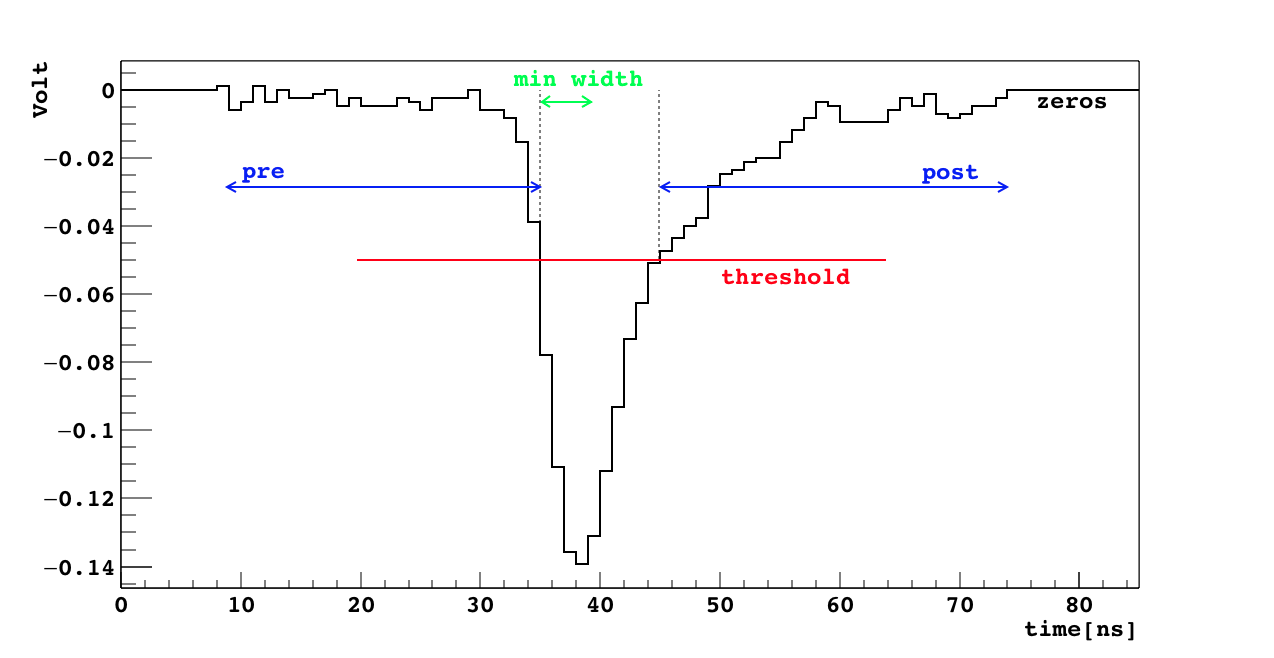} 
\caption{Zero suppressed waveform of a single photoelectron pulse.
The red line represents the zero suppression threshold in amplitude (set to $-0.5$\,mV in this example).
The green line represents the minimum width, which refers to the minimum number of samples the waveform must stay above the threshold.
The blue lines represent pre- and post-samples.}
\label{fig:zspulse}
\end{center}
\end{figure}

Each \pxie\ waveform digitizer module asynchronously and continuously samples the signals coming from the $\times$10 outputs of the FE modules.
When a trigger is received, a block of samples around the trigger (the \emph{acquisition window}) is stored in the digitizer memory buffer.
The position of the trigger inside the acquisition window (the \emph{trigger reference position}) is configurable, such that samples prior to the trigger can be stored. 
This allows the system to compensate for delays due to signal propagation and trigger generation.
The acquisition windows and trigger reference positions used for data acquisiton are described in section~\ref{sec:veto_daq_trigger}.

The trigger signal is received in the \pxie\ timing and synchronization module present in each \pxie\ chassis, to route the trigger condition to every \pxie\ digitizer in the \pxie\ chassis using the \pxie\ bus.
This timing module also receives an external $50$\,MHz clock, the 1PPS signal, the run enable signal and the serial encoded \trgID\ signal.
The $50$\,MHz clock is used to synchronize the devices handled in the \pxie\ chassis to the main clock of the DarkSide-50 experiment.
The digital lines are routed to the \pxie\ \flex\ \fpga\ module that decodes the \trgID\ and generates the timestamps to uniquely identify the event and correlate it with TPC triggers. 
More details on the \fpga\ logic, \trgID\ and timestamps are described in section~\ref{sec:veto_daq_trigger}.

 Sample waveform data recorded by the \pxie\ digitizers are fetched asynchronously by the readout software \odaq\, running on the \pxie\ controller.
\odaq\ waits for a configurable number of waveform data blocks (\emph{number of records to fetch}) to be stored in the \pxie\ digitizer memory and readied to be fetched.
When this condition is met, \odaq\ starts fetching the records.

To reduce the size of waveform data, a zero-suppression algorithm is applied in \odaq. 
The zero-suppression algorithm is described in section~\ref{sec:zs}.
Zero-suppressed data from the different \pxie\ digitizers in the same \pxie\ chassis are then bundled in a data fragment structure, which also contains the timestamp and the \trgID\ of the event.
Data fragments are then transmitted to \odb\ via TCP/IP, using the simple messaging reference library (STM)~\cite{STM}.

During data acquisition, \odb\ collects data fragments from every \pxie\ chassis.
For each trigger, the consistency of \trgIDs\ and timestamps of data fragments received from different \pxie\ chassis is checked, and a unique raw event with the pulses collected by all channels is written to disk.
The raw file of the veto is a custom binary file which contains a header with all the information on the hardware and software configuration of the data acquisition, which is followed by the timestamp, \trgID, and the zero-suppressed waveforms of each event.

\odb\ provides also an online monitor and event display.
The trigger rate and the data transfer rate for each chassis are displayed.
The user can also display, for the \lsv\ and \wcd, histograms of the number of PMTs with at least one pulse, the total number of pulses in a trigger,
a waveform graph for a selected channel, and the summed waveform over all channels of the \lsv\ and \wcd.

\subsection{Zero-suppression} 
\label{sec:zs}

The zero-suppression algorithm lets the user specify a threshold in amplitude, minimum width, and  number of pre- and post-samples.
When the waveform crosses the threshold and stays above threshold for the number of samples specified by the minimum width, the algorithm returns the entire waveform between both threshold crossings, including also the specified number of pre-samples before the first crossing and the specified number of post-samples after the final crossing. 
If the waveform goes below  threshold and then back above before the specified number of post-samples have passed, the algorithm waits for the waveform to drop back below threshold and then starts counting post-samples starting from zero. 
Therefore, if two pulses appear on the same channel with overlapping zero-suppression windows, they are combined into one larger zero-suppressed pulse.
The output of the zero-suppression algorithm is called the zero-suppressed waveform (or \emph{pulse}, or zero-suppressed pulse).
All the samples outside the zero-suppressed pulses are discarded.

During the first run of \dsf, veto data were usually taken with a zero-suppression threshold of $-30$\,mV (corresponding to $\sim0.25$\,PE), a minimum width of 3 samples ($2.4$\,ns), and $20$\,ns worth of pre-samples and post-samples.
Figure~\ref{fig:zspulse} shows a zero-suppressed waveform of a single photoelectron signal. 

\subsection{Veto trigger} 
\label{sec:veto_daq_trigger}

There are 2 independent configurations for the input of triggers into the veto \daq:
\begin{itemize}
	\item \emph{\tpc\ trigger mode}: in this configuration the trigger signal of the veto is delivered only by the \tpc\ \daq. 
The length of the data acquisition window is usually set in a range from $70$ to $200$\,$\mu$s in order to detect prompt and delayed physical events in the veto correlated to the \lar\ \tpc;
	\item \emph{Veto self-triggering mode}: In this configuration the input trigger of the \daq\ is the logical ``or'' between the \lsv\ and \wcd\ triggers delivered by the \fedb\ (see Section~\ref{sec:fedb_trg}). 
The trigger rate is dominated by the \lsv\ trigger rate and depends strongly on the \lsv\ majority threshold. 
The length of the data acquisition window is usually set to $6.5$\,$\mu$s in order to acquire the whole scintillation signal, which has a duration of about $300$\,ns plus electronics overshoot, undershoot and afterpulses that naturally follows events with more than $\sim10$\,PE per channel.
\end{itemize}

The data used for the \wimp\ search are taken in \tpc\ trigger mode.

In the first phase of \dsf, with a 50\% mixture of \pc\ and \tmb\ in the \lsv, the acquisition gate was set to $70$\,$\mu$s, of which $10.5$\,$\mu$s came prior to the trigger, and $59.5$\,$\mu$s after the trigger (corresponding to a reference position of $15$\%).

In the second phase of \dsf, with mixture of \pc\  (95\%) and \tmb\ (5\%) in the \lsv, the acquisition gate was set to $140$\,$\mu$s, of which $10.5$\,$\mu$s came prior to the trigger and $129.5$\,$\mu$s after the trigger. 
The acquisition gate was later extended to $200$\,$\mu$s, of which $10.5$\,$\mu$s came prior to the trigger and $189.5$\,$\mu$s after the trigger.

The data acquired during the calibration campaigns of \dsf\ were taken using both triggering modes.
Information about the \daq\ of the \lar\ \tpc\ of \dsf\ can be found in~\cite{tpcdaq}.

The \lar\ \tpc\ and veto systems can trigger independently, but a synchronization mechanism must be provided in order to correlate the \lar\ \tpc\ events with veto signals. 
The two \daq\ systems are physically displaced, therefore synchronization between the two systems is obtained through a high precision time-stamp obtained from a common $50$\,MHz clock slaved to a 1PPS signal received from the LNGS GPS.

The trigger signal is delivered to each of the 4 \pxie\ timing modules and distributed to all the devices in the \pxie\ chassis through an internal bus.
The \pxie\ timing modules also receive the $50$\,MHz clock, the 1PPS signal, and the run enable signal from the \tpc\ \daq. 
These signals are used to compute the timestamp of the event using three counters:
\begin{itemize}
\item the 1PPS counter which counts the number of seconds elapsed since the run enable signal was delivered (which correspond to the start of the run);
\item the GPS fine time counter which counts the number of $50$\,MHz clock cycles elapsed since the run enable or the 1PPS signal (whichever comes last);
\item the GPS one second counter which counts the number of $50$\,MHz clock cycles elapsed since the most recent 1PPS signal.
\end{itemize}

The three counters are evaluated for every trigger received, then stored in a DMA FIFO of the \pxie\ \flex\ \fpga and read asynchronously by \odaq.
These counters allow the system to compute the time of the trigger since the start of the run with a precision of $20$\,ns.
The same timestamp is also computed in the \tpc\ \daq\ system, allowing physical events in the \lar\ \tpc\ and the vetoes to be correlated with each other, even in veto self triggering mode.

In the \tpc\ trigger mode, a 16-bit \trgID\ is generated for each trigger by the \tpc\ \daq\ and serially encoded using a frequency modulation scheme.
The \trgID\ signal is acquired by the \pxie\ timing module and routed to the \pxie\ \flex\ \fpga\ where it id decoded.

\subsection{Communication and synchronization with the \lar\ \tpc\ \daq} 
\label{sec:comm_tpc}

\begin{figure}[tbp]
\begin{center}
\includegraphics[width=0.7\columnwidth]{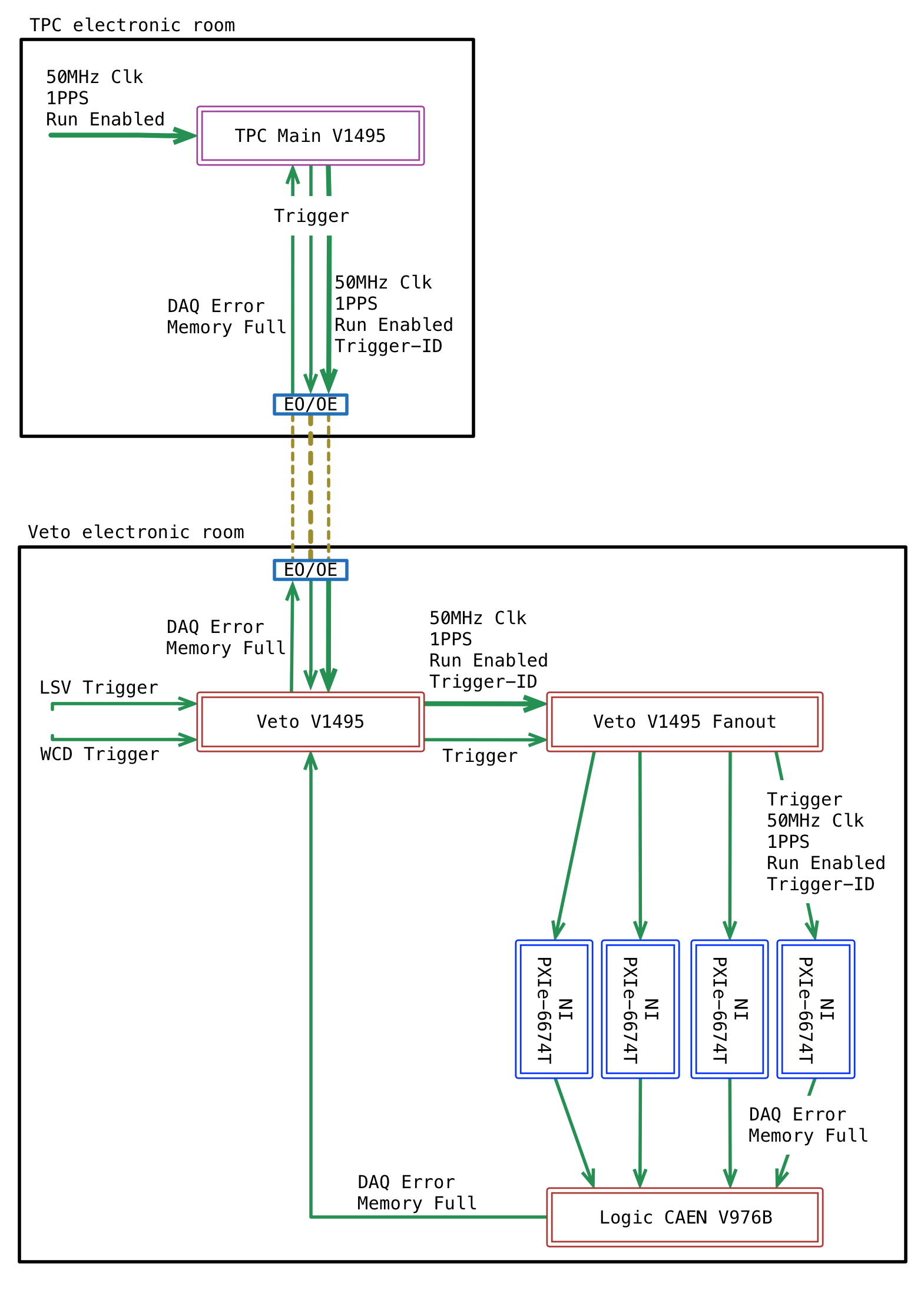} 
\caption{A block diagram showing the trigger connections between the \tpc\ \daq\ and the veto \daq. 
EO means electrical to optical conversion.}
\label{fig:comm_tpc}
\end{center}
\end{figure}

The scheme in figure~\ref{fig:comm_tpc} describes the hardware communication between the \tpc\ \daq\ and the veto \daq.
Unless otherwise noted, signals in the block diagram are active high TTL, ECL or LVDS. 
The hardware connections between the \tpc\ \daq\ and the veto \daq\ are based on optical links and \fpga\ units~\cite{NI7961, caenV1495}.
The electrical-optical converters are VME V730-13 and V720-13 modules.
The \fpga\ units are:
\begin{enumerate}
\item the \tpc\ V1495 main trigger module located at the \tpc\ electronics room;
\item a V1495 trigger module located at the veto electronics room;
\item a fanout V1495 module located at the veto electronics room;
\item the veto \pxie\ \flex\ \fpga\ logic modules located in the \pxie\ crates, described above.
\end{enumerate}

In what follows we focus on the communication between the \tpc\ \daq\ and triggering with the veto system.
The details of the \tpc\ \daq\ and trigger are described in~\cite{tpcdaq}.

The \tpc\ V1495 main trigger module, located at the \tpc\ electronics room, generates the \tpc\ triggers necessary to trigger the veto when running in \tpc\ trigger mode.
The trigger signal, the run enable signal, and a 16-bit \trgID\ are transferred via optical link to the veto V1495 trigger module located at the veto electronics room.
A $50$\,MHz clock signal is also sent from the \tpc\ electronics room to the veto electronics room for synchronization of all veto and \tpc\ \daq\ devices to the same clock.
The V1495s and the veto \pxie-6674T module also receive the 1PPS signal.
The logic inside veto V1495 trigger module decodes the Trigger ID and places it in an internal FIFO memory for the readout via the VME bus as a part of the \tpc\ data stream.
All VME FPGAs and electrical-optical converters are allocated in a VME crate.
 
Four copies of the trigger signal, the run enable signal, the \trgID, the common $50$\,MHz clock, and the 1PPS signal are generated by the fanout V1495 module and delivered to each veto \pxie-6674T module in the four \pxie\ chassis.
The \pxie-6674T handles the synchronization of any clock inside the \pxie\ chassis to the common $50$\,MHz clock, allowing the synchronization of all veto digitizers. 
The \pxie-6674T also routes the trigger signal to the digitizers through the \pxie\ bus, allowing for a synchronous trigger among all devices.
Finally, the \pxie-6674T routes the trigger signal, the run enabled signal, the serially encoded \trgID, and the 1PPS signal to the \pxie\ \flex\ \fpga\ modules, which decodes the \trgID\ and generates the timestamp, as described in Section~\ref{sec:veto_daq_trigger}.
The \pxie\ \flex\ \fpga\ module has internal DMA memory which is used as a FIFO to store \trgID\ and timestamp counters for each incoming trigger.
This memory is read out with each event in order to provide trigger markers for \tpc\ and veto synchronization.

A Memory Full and a DAQ Error signal can be produced by each of the \pxie-6674T modules in the veto \daq\ system in order to keep data readout rate at the maximum.
This signal is sent to the \tpc\ electronics to prevent generation of the \tpc\ trigger (and associated veto triggers) when the veto \daq\ is close to filling all available memory buffers.
The logical ``or'' of the four Memory Full and DAQ Error signals is performed by a CAEN V976B logic module.

\subsection{Veto TDC} \label{sec:tdc}

A secondary data stream for the veto system is implemented using TDCs. 
As described in Section~\ref{sec:fe}, one of the discriminated \fedb\ veto outputs from each channel is connected to a CAEN V1190 multi-hit 128 channel TDC~\cite{caen1190}. 
The two V1190 modules are located in the veto electronics room and connect to the \tpc\ V1495 trigger unit through the veto V1495 fanout logic unit. 
The trigger signal for the TDC modules is  generated by the veto V1495 trigger module.
The TDC \daq\ software is built within the Fermi National Accelerator Laboratory's (Fermilab) {\it art} framework~\cite{art}.

This secondary system does not replace the main veto readout but provides a back-up system.
TDCs are only triggered by the \tpc, even when the veto runs in self trigger mode, therefore each \tpc\ trigger has direct mapping to the veto TDC data. Consequently, the TDC provides an independent data stream for the \lsv\ and \wcd, and is synchronized to the \tpc\ trigger providing a cross-check on the efficiency of the ADC-based veto \daq.

\subsection{Run control} \label{sec:rc}

The veto \daq\ and \tpc\ \daq\ systems are handled by a common run controller which is configured to permit different data acquisition modes:
\begin{itemize}
\item global runs, where the \tpc\ and veto run at the same time (choosing either \tpc\ trigger mode or veto self-trigger mode) and
\item local runs, where the sub-detectors run independently.
\end{itemize}

The run controller application has been developed in \labview, and conforms to \labview\ coding and documentation standards. 
The run controller handles the communication to all \daq\ devices and processes.
Communication over ethernet with the \tpc\ sub-systems is performed using the XML-RPC protocol, while communication with veto sub-systems is performed using the \labview\ STM protocol.
The run controller application has a graphical user interface that allows an operator to initialise or reset \daq\ configurations, start and stop the data acqusition.
The run controller supervises the data acquisition. When either the \tpc\ or the veto \daq\ display unusual behaviour or errors,  data acquisition is automatically stopped. All \daq\ sub-systems are then reset and re-initialized, and new run is automatically started.
The run controller also logs the relevant parameters of each run to the experiment database.

%% file: performances.tex
\section{Performance} 
\label{sec:performances}

\begin{figure}[tbp]
\begin{center}
\includegraphics[width=0.85\columnwidth]{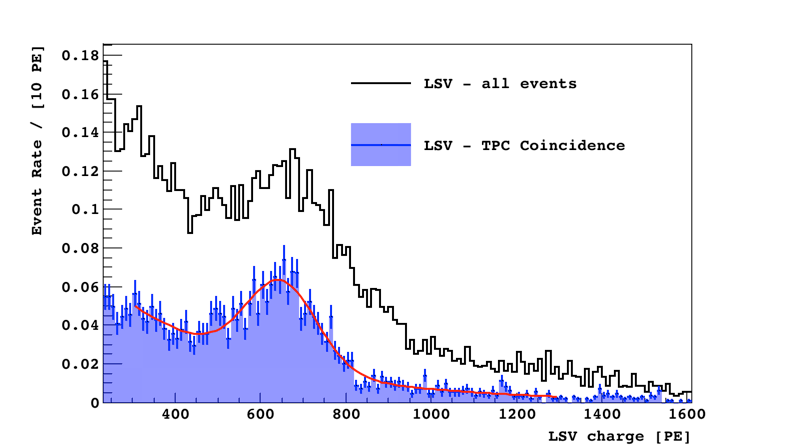}
\caption{Charge spectra of a sample of events is the \lsv\ collected during the first dark matter run, in \tpc\ trigger mode, with an acquisition gate of $70$\,$\mu$s. 
The black line is the spectrum of all events, the blue filled histogram is the spectrum in a window of $100$\,ns in coincidence with the \tpc. 
The peak is due to one of the two \gr\ emitted in the decay of \coba. 
The resolution of the \lsv\ is not enough to separate the energy of the two \grs, which are $1.17$\,MeV and $1.33$\,MeV. 
More informations on the reconstruction and analysis algorithms used to generate these spectra can be found in~\cite{ds50first, ds50veto}.}
\label{fig:co60}
\end{center}
\end{figure}

\begin{figure}[tbp]
\begin{center}
\begin{tabular}{cc}
\includegraphics[width=0.5\columnwidth]{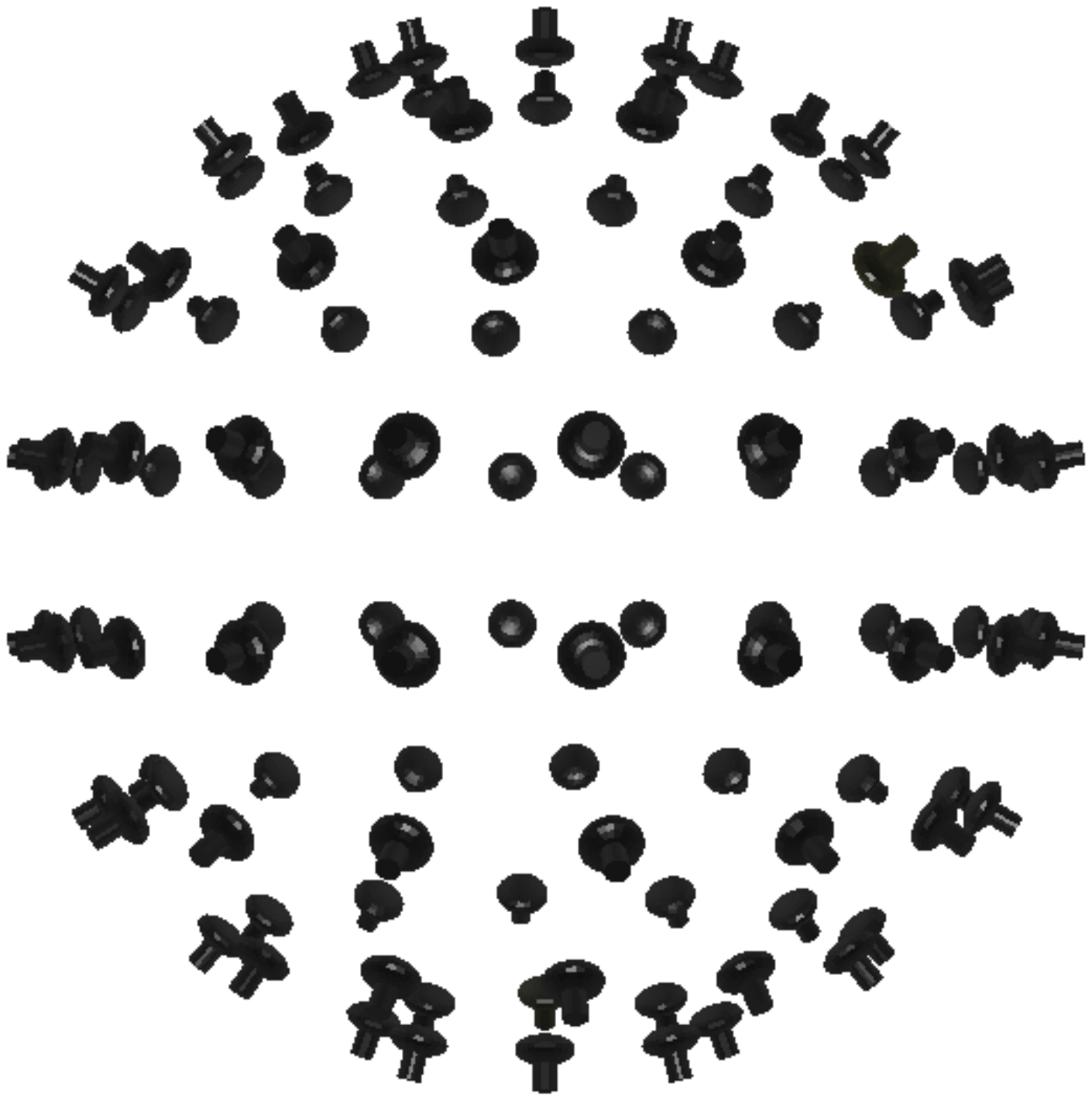}&
\includegraphics[width=0.5\columnwidth]{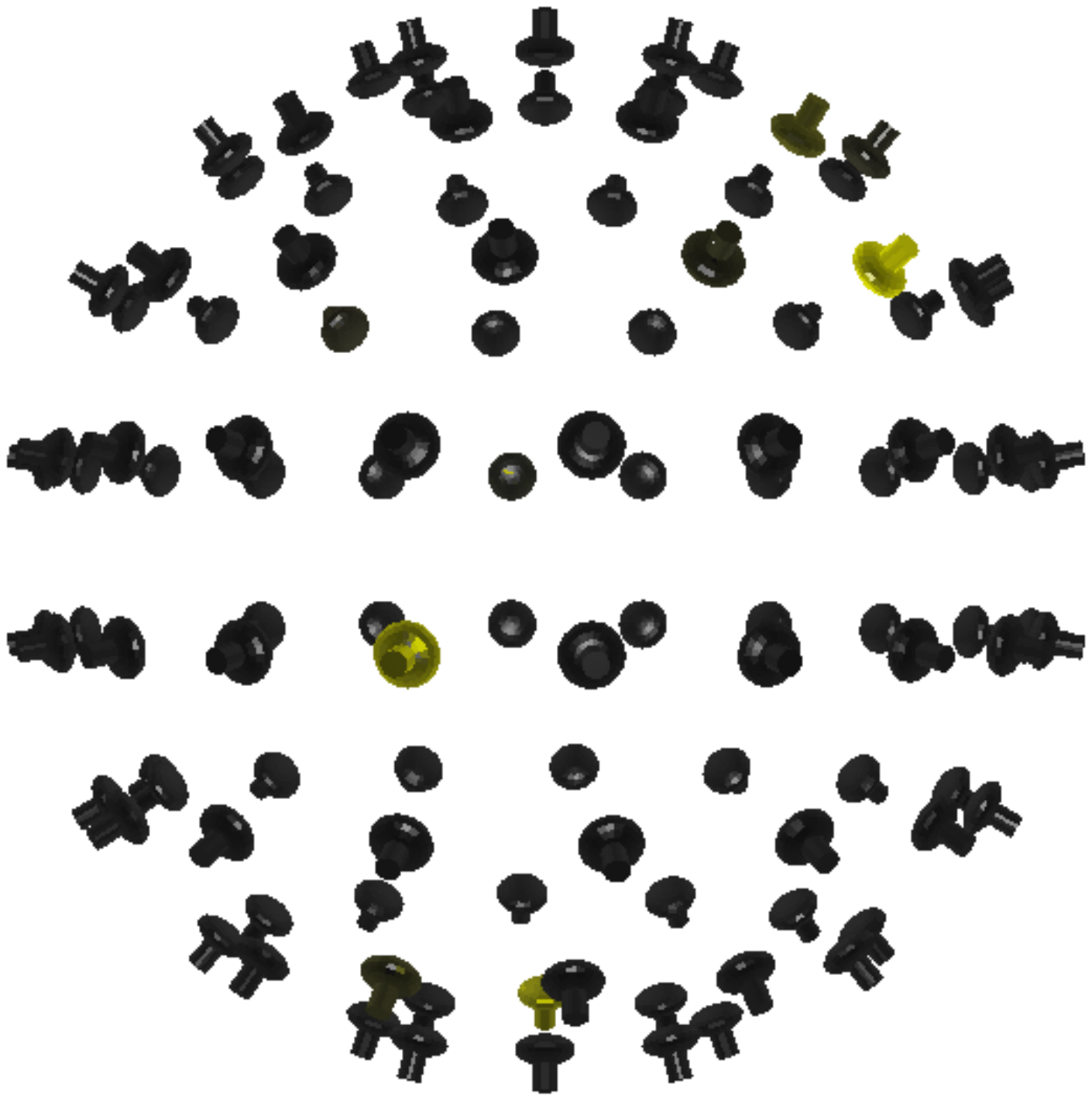}\\
\includegraphics[width=0.5\columnwidth]{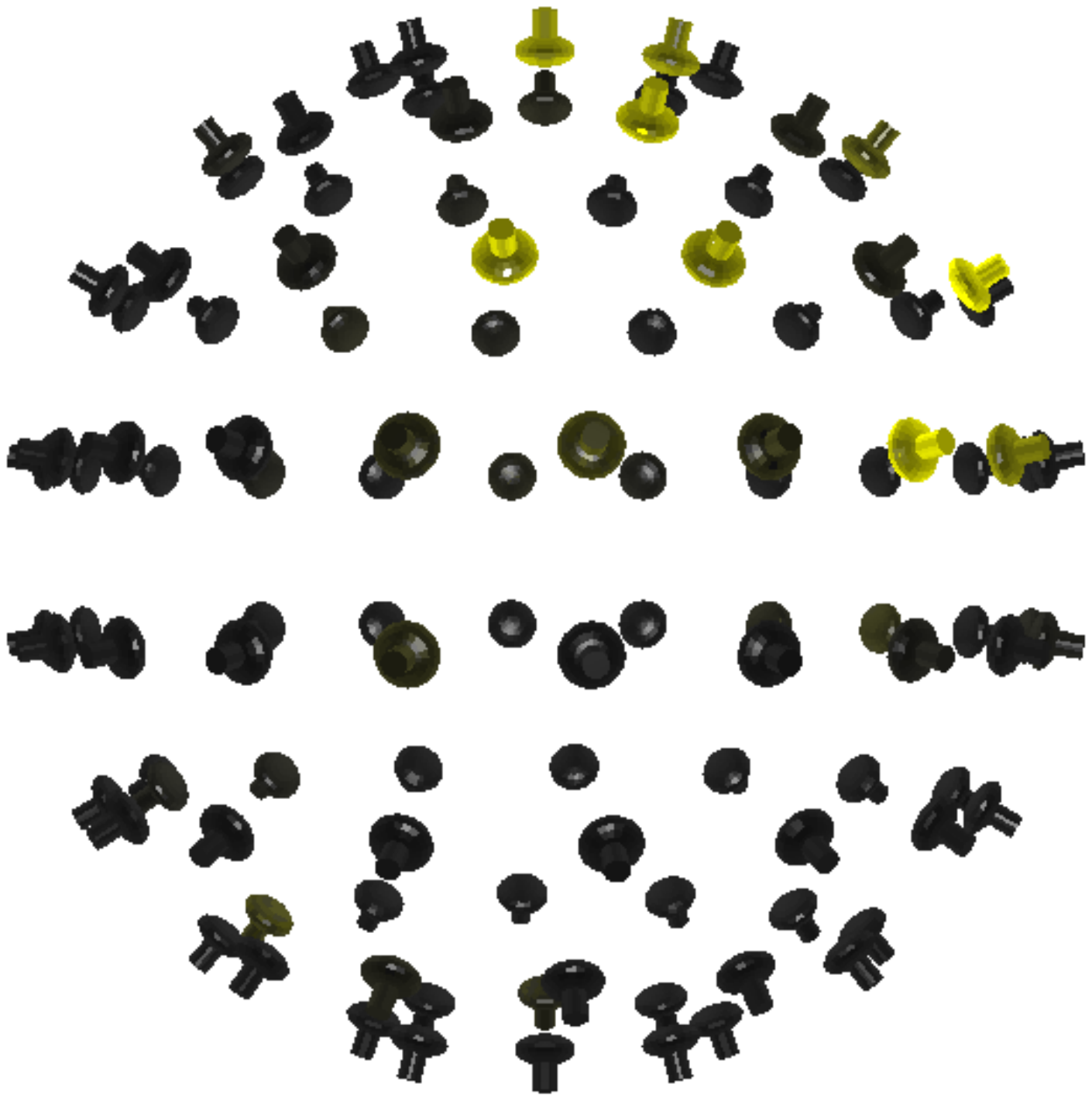}&
\includegraphics[width=0.5\columnwidth]{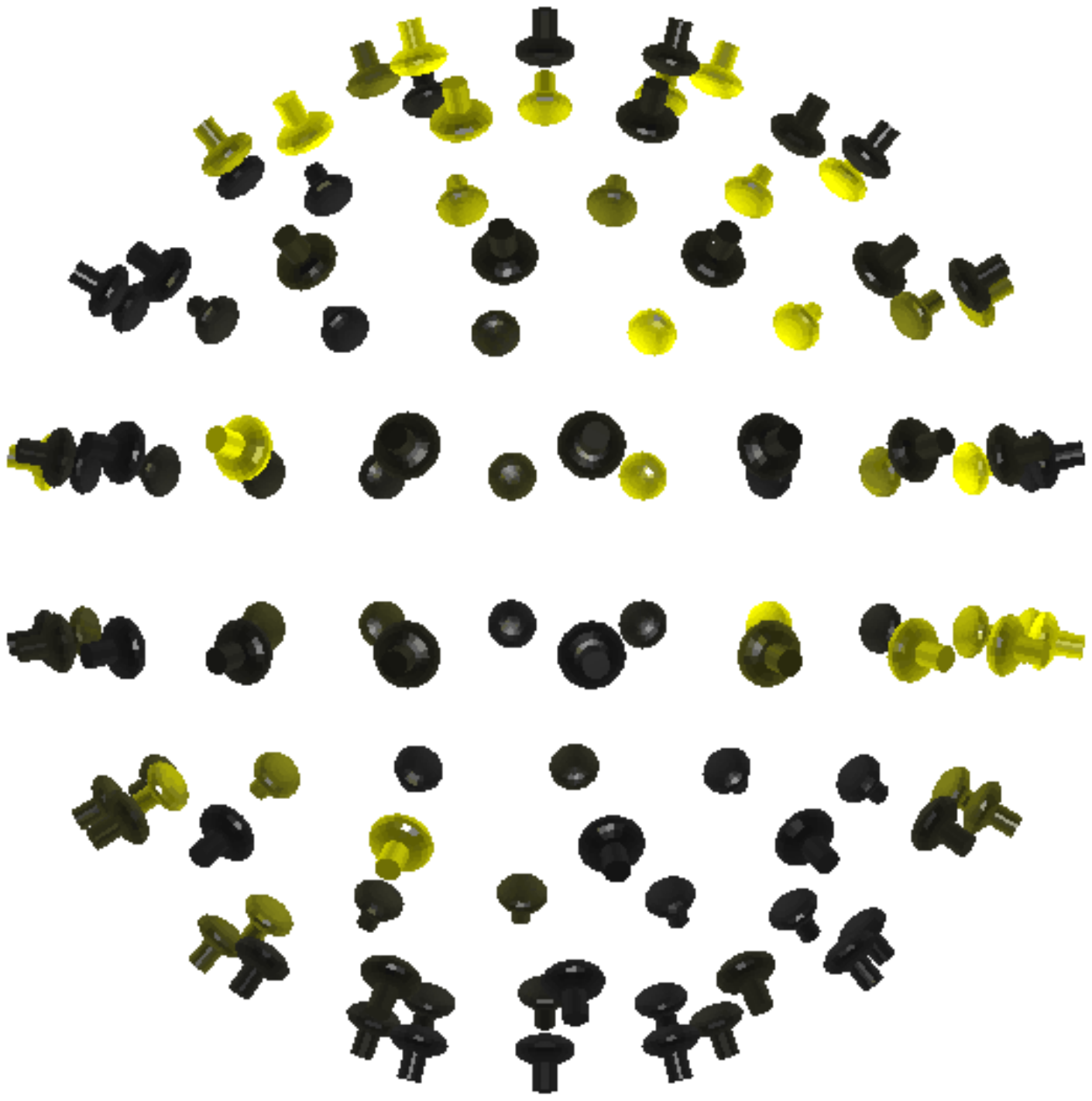}\\
\includegraphics[width=0.5\columnwidth]{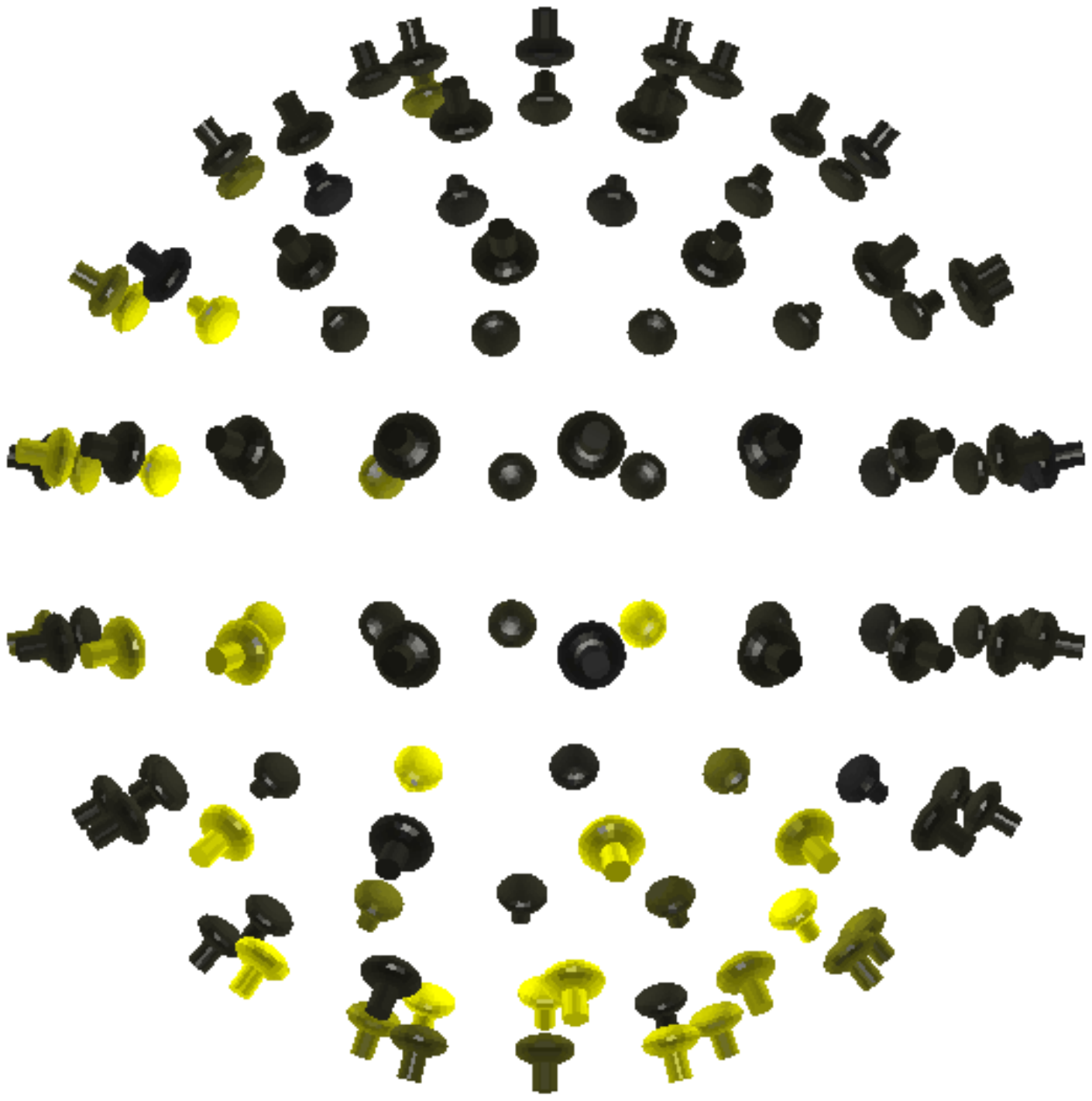}&
\includegraphics[width=0.5\columnwidth]{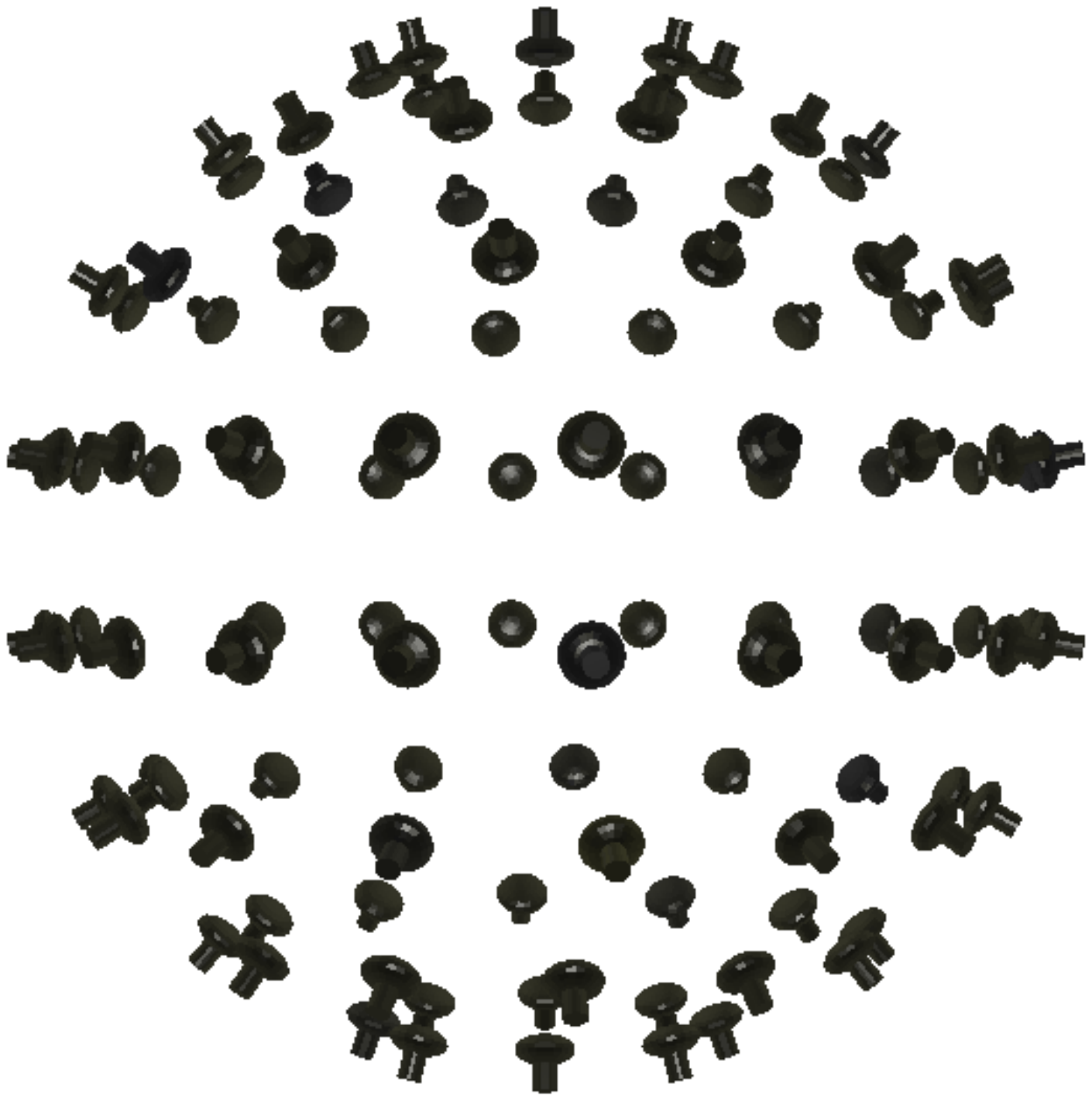}
\end{tabular}
\caption{Event display of a muon crossing the \lsv.
The intensity of the color on the \pmt\ is proportional to the waveform amplitude. 
The frames are sorted from then top to bottom.
Each of the 6 frames is delayed of $8$\,ns with respect to the previous frame.
More informations on the reconstruction algorithms used to generate this event display can be found in~\cite{ds50first, ds50veto}.}
\label{fig:muon}
\end{center}
\end{figure}

\dsf\ began its first physics run in November 2013, with the veto electronics and \daq\ system described above.
At that time, the \lsv\ showed an unexpectedly high activity of \cfor\ (\lsvcforrate) in the \tmb, whose feedstock was partly derived from modern carbon that has a much higher \cfor\ content than petroleum-derived~\cite{ds50first, ds50veto}.
Despite the fact that during this initial period of data tacking the \cfor\ activity was a factor $\sim500$ higher than design goals, the veto \daq\ was able to sustain the data throughput, allowing a duty cycle higher than 95\% during \wimp\ search runs.
The veto  electronics and \daq\ system has also been used to acquire data during calibration campaigns with neutron and $\gamma$ radioactive sources deployed in the \lsv, and it is now being used for the \wimp\ search run with a cleaner \lsv\ and the \lar\ \tpc\ filled with \uar.
The full \daq\ system has been operational since November 2013 and can be remotely controlled.

The veto data acquisition is stable with a maximum input data throughput of about $1$\,GByte per second on each \pxie\ chassis (each of the two \lsv\ chassis acquire 55 channels and each of the two \wcd\ chassis acquire 40 channels).
The trigger rate that the system can sustain depends on the the length of the acquisition window and the activity of the detector.
After the first \wimp\ search run, the \cfor\ activity in the \lsv\ was significantly reduced, from $150$\,kBq to $0.3$\,kBq, allowing longer acquisition windows for each \tpc\ trigger. 
In the current detector condition, the maximum trigger rate sustainable with an acquisition window of $200$\,$\mu$s is $\sim16$\,Hz in \tpc\ triggering mode.
The output data rate in the current \wimp\ search mode, using an acquisition window of $200$\,$\mu$s, is $\sim19$\,kByte for a complete \lsv\ and \wcd\ event.
With a usual \lar\ \tpc\ trigger rate of $\sim1.5$\,Hz~\cite{ds50second}, the veto \daq\ writes $\sim2.5$\,GByte to disk per day (before bzip2 compression that decreases the size of the data by about a factor of 4).
The \daq\ system is constantly improved in performance, to maximize stability and introduce new features stemming from the requests from the experiment.

The synchronization between the \lar\ \tpc\ and the \lsv\ is checked and monitored using physical events that generate scintillation light at approximately the same time in both detectors, such as muons or \coba\ decays.
\coba, which can be found in the \tpc\ cryostat, decays with emission of two \grs. 
The geometry of \dsf\ allows for one of the $\gamma$-rays to reach, and trigger, the \lar\ \tpc, while the other produces scintillation light in the \lsv.
These events are used as benchmark to define the \emph{prompt coincidence region} between the \lar\ \tpc\ and the veto.
A sample of the charge spectrum of the events in the \lsv\ is displayed in Figure~\ref{fig:co60}, where the peak due to \coba\ decay in a narrow time window coincidence with the \lar\ \tpc\ scintillation signal is clearly visible.

The synchronization between the \lsv\ and the \wcd\ is checked and monitored using muon events which cross both detectors.
The event display of a muon crossing the \lsv\ and is displayed in Figure~\ref{fig:muon}.

%% file: conclusions.tex
\section{Conclusions}

This paper reports the description of the electronics and \daq\ system developed for the veto detectors of the \dsf\ experiment. 
The \daq\ involves data collection from a liquid scintillator veto and a water Cherenkov detector.

The front end electronics, the \daq, and the trigger system discussed here have been used to acquire data in the form of zero-suppressed waveform samples from of the 110 \pmts\ of the \lsv\ and the 80 \pmts\ of the \wcd.
The synchronization between the veto \daq\ and the \tpc\ \daq\ is also described.

The electronics and \daq\ have been used since the start of the \dsf\ experimental phase to acquire data for dark matter runs and calibration campaigns with radioactive sources.
It is foreseen to use the electronics and \daq\ system described here throughout the remainder of \dsf\ operations.
The veto \daq\ system has proven its performance and reliability even with an event rate $\sim500$ time higher than the design goal.
The Veto \daq\ system has ensured stable data taking and allowed for a high duty cycle of data taking.

The \daq\ described here is entirely based on \labview. Any programming language could have been used to write the code, but the \labview\ approach made it possible for a small team to quickly develop a high-throughput, fast-digitizer \daq\ system with a large number of channels.

Because of the demonstrated performance and reliability, the whole electronics and \daq\ system architecture of the \dsf\ veto detectors is scalable for the veto detectors of a multi-ton DarkSide experiment, such as the proposed DarkSide-20k detector.